\documentclass{aa}  
\usepackage{txfonts}
\usepackage{epstopdf}

\usepackage{graphicx,longtable,lscape,natbib,amssymb,amssymb,amsmath}
\bibpunct{(}{)}{;}{a}{}{,} 
\newcommand{\ltsima} {$\; \buildrel < \over \sim \;$}  
\newcommand{\gtsima} {$\; \buildrel > \over \sim \;$}  
\newcommand{\lta} {\lower.5ex\hbox{\ltsima}}  
\newcommand{\gta} {\lower.5ex\hbox{\gtsima}}  
\newcommand{\Ha} {H$\alpha$}

\newcommand{\kms}{$\rm{\,km \,s}^{-1}$}

\newcommand{\forb}[2]{\mbox{$[{\rm #1\, #2}]$}}
\newcommand{\oiii}{\forb{O}{III}}
\newcommand{\oi}{\forb{O}{I}\,}
\newcommand{\sii}{\forb{S}{II}\,}
\newcommand{\nii}{\forb{N}{II}\,}

\usepackage{color}

\begin{document}

\title{The HST view of the innermost narrow line region}
\subtitle{} \titlerunning{The Innermost NLR} 
\authorrunning{B. Balmaverde et al.}

\author{Barbara Balmaverde\inst{1,2}
\and Alessandro Capetti\inst{3}
\and Daria Moisio\inst{3}
\and Ranieri D. Baldi\inst{4,5}
\and Alessandro Marconi\inst{1,2}}

\institute {Dipartimento di Fisica e
  Astronomia, Università di Firenze, via G. Sansone 1, 50019 Sesto
  Fiorentino (Firenze), Italy
\and INAF - Osservatorio Astrofisico di Arcetri, Largo Enrico
  Fermi 5, I-50125 Firenze, Italy
\and INAF - Osservatorio Astrofisico di Torino, Via Osservatorio 20,
  I-10025 Pino Torinese, Italy
\and Department of Physics, Technion, 3200, Haifa, Israel
\and Department of Physics and Astronomy, The University, Southampton SO17 1BJ, UK}

\offprints{balmaverd@arcetri.inaf.it} 

\abstract{We analyze the properties of the innermost narrow line region in a
  sample of low-luminosity AGN. We select 33 LINERs (bona fide AGN) and
  Seyfert galaxies from the optical spectroscopic Palomar survey observed by
  HST/STIS. We find that in LINERs the \nii and \oi lines are broader than the
  \sii line and that the \nii/\sii flux ratio increases when moving from
  ground-based to HST spectra. This effect is more pronounced considering the
  wings of the lines.  Our interpretation is that, as a result of superior HST
  spatial resolution, we isolate a compact region of dense ionized gas in
  LINERs, located at a typical distance of $\sim$3 pc and with a gas density
  of $\sim$10$^4$--10$^5$ cm$^{-3}$, which we identify with the outer portion
  of the intermediate line region (ILR). Instead, we do not observe these
  kinds of effects in Seyferts; this may be the result of a stronger dilution
  from the NLR emission, since the HST slit maps a larger region in these
  sources. Alternatively, we argue that the innermost, higher density
  component of the ILR is only present in Seyferts, while it is truncated at
  larger radii because of the presence of the circumnuclear torus. The ILR is
  only visible in its entirety in LINERs because the obscuring torus is not
  present in these sources.}

\maketitle

\section{Introduction}
\label{intro}

The narrow line region (NLR) is a unique laboratory to explore the properties
of the interstellar medium at the boundary between the host galaxy and the
region dominated by the active galactic nucleus (AGN). This is the region
where most of the energy exchange between the AGN and its host galaxies
occurs. In the nearby Universe, many strong and prominent narrow emission
lines are accessible in the optical bands.  For these reasons, the properties
of the NLR has been extensively explored since the 80s with ground-based
spectra. For example, the spectroscopic survey of bright nearby galaxies
\citep{filippenko85,ho95}, hereafter the Palomar survey, provides a unique
catalog of optical spectra and narrow emission line measurements.  However,
as a result of the superior spatial resolution of HST, we have the opportunity to
explore the properties of the innermost narrow line region, unaccessible by
the ground-based observations. The HST might provide us with a different view,
particularly for nearby, low-luminosity AGN with respect to what is seen at
larger scales.

In the early literature there are disagreements between different
studies. This is likely because of the small size of the samples and/or
low-resolution effects. In particular, it is not clear how properties of the
emitting gas (ionization level, temperature, density, reddening, velocities,
etc.) are linked to the characteristics of the AGN, such as its luminosity or
spectral type. Evidence for both positive correlations between line width and
ionization/critical density of the lines are observed in a number of LINERs
\citep{filippenko85} or Seyferts \citep{veilleux91, moore96}; in other cases a
negative correlation emerges (e.g., \citealt{veilleux91}). These studies
indicate that a relation between the density and/or ionization parameter
exists with velocity, at least in some objects. Since the velocity increases
toward the center, these observed correlations imply that the density and/or
ionization parameter gradually increases inward in the NLR of these galaxies.
However, the main results about the correlations of line width (and also on
the presence of overall line blueshift) with ionization potential and/or
critical density are based on emission lines of high ionization
([Fe~VII]$\lambda6087$, [Fe~X]$\lambda6374$, [Ne III]$\lambda$3869; e.g.,
\citealt{marconi96}) but no variation in width is found between, for example,
\oi and \sii (see, for example, Fig. 3b in \citealt{moore94}).

With the advent of HST, a structure characterized by emission line velocities
intermediate between those from the BLR and NLR has been identified in
Seyferts (e.g., \citealt{crenshaw07,crenshaw09}). This region has been called
intermediate line region (ILR). Assuming that the velocity widths of the ILR
is primarily due to gravitational motion, the distance from the ILR to the
central ionizing source is $\sim$ 0.1-1 pc.  In this region, the different
velocities of the lines can be explained as the result of outflowing gas or a
stratification in the ionization state and/or density of the gas
\citep{kraemer00}.

The importance of the NLR density stratification has been demonstrated, e.g.,
for its impact on our ability to isolate the BLR emission in low-luminosity
AGN. A widely used strategy consists in using the [S II] doublet as a template
to subtract the contribution of the narrow emission lines to the
H$\alpha$+[N~II] complex.  In HST spectra, however, the [O~I] line is generally
broader than the  [S~II] line and using the [O~I] doublet instead of [S~II] leads to
completely different results.  We have shown in \citet{balmaverde14} that this
approach does not lead, in general, to robust constraints on the properties of
the BLR in low-luminosity AGN.

In this paper we focus on the view of the NLR by ground-based and HST/STIS
spectra, analyzing the profile parameters of low ionization emission lines
(\oi$\lambda\lambda$6300,6363, \nii $\lambda\lambda$6548,6584, \Ha, and [S
  II]$\lambda\lambda$6716,6731). We take advantage of the high spatial
resolution of HST to deal with the still debated topic of the gas density
stratification of the NLR and on the relation between the emission line width
and the critical density of the gas in nearby AGN.

The paper is organized as follows. In Sect. \ref{sample} we present the sample
and methodology of spectral analysis. In Sect. 3 we compare the properties
of the emission line profile in HST and Palomar spectra, then we present the
main result for LINERs (Sect. 4) and Seyferts (Sect. 5). In Sect. 6 and 7 we
provide a discussion,  our summary, and conclusions.

\begin{figure}
\centering{ \includegraphics[width=9cm,angle=0]{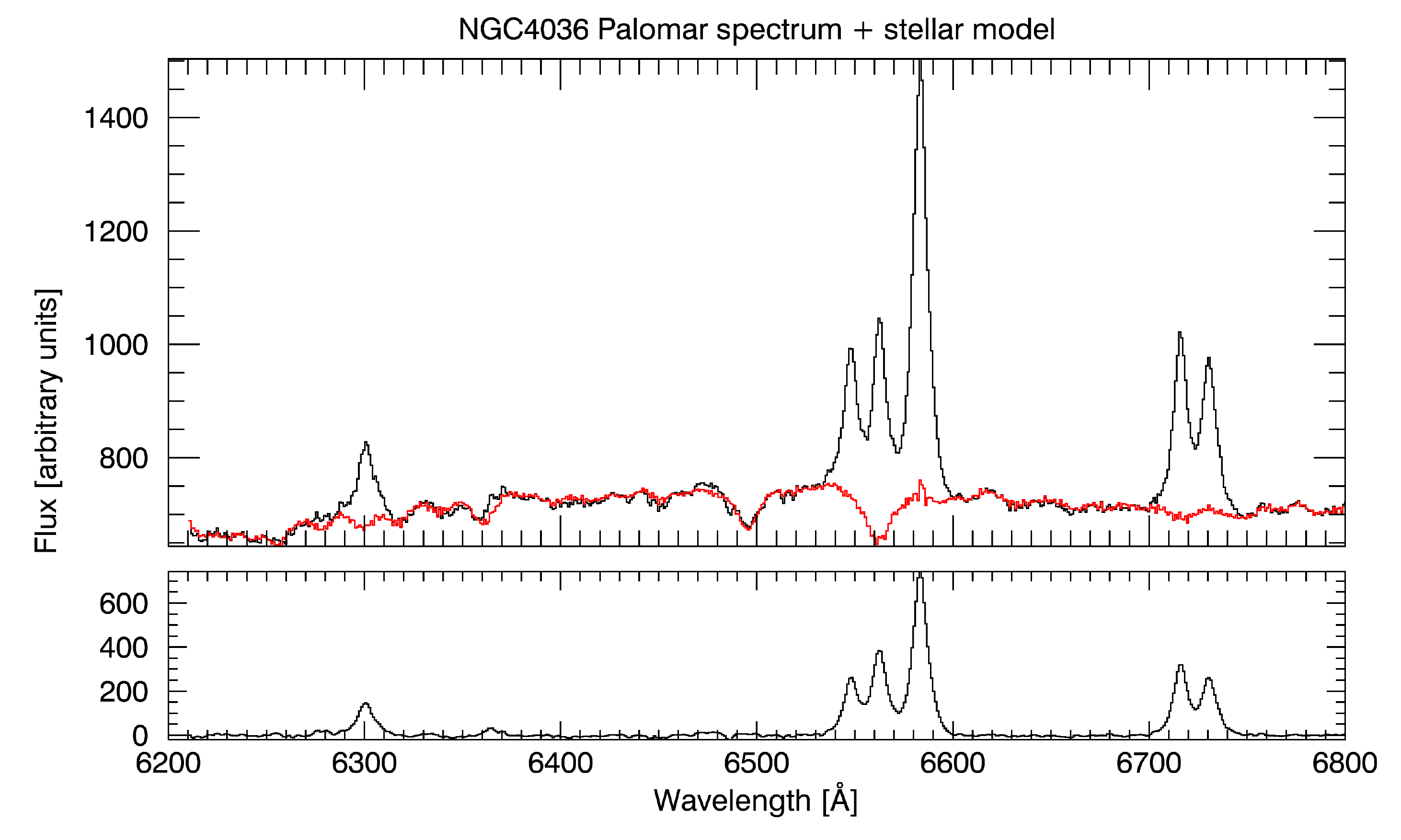}
  \includegraphics[width=9cm,angle=0]{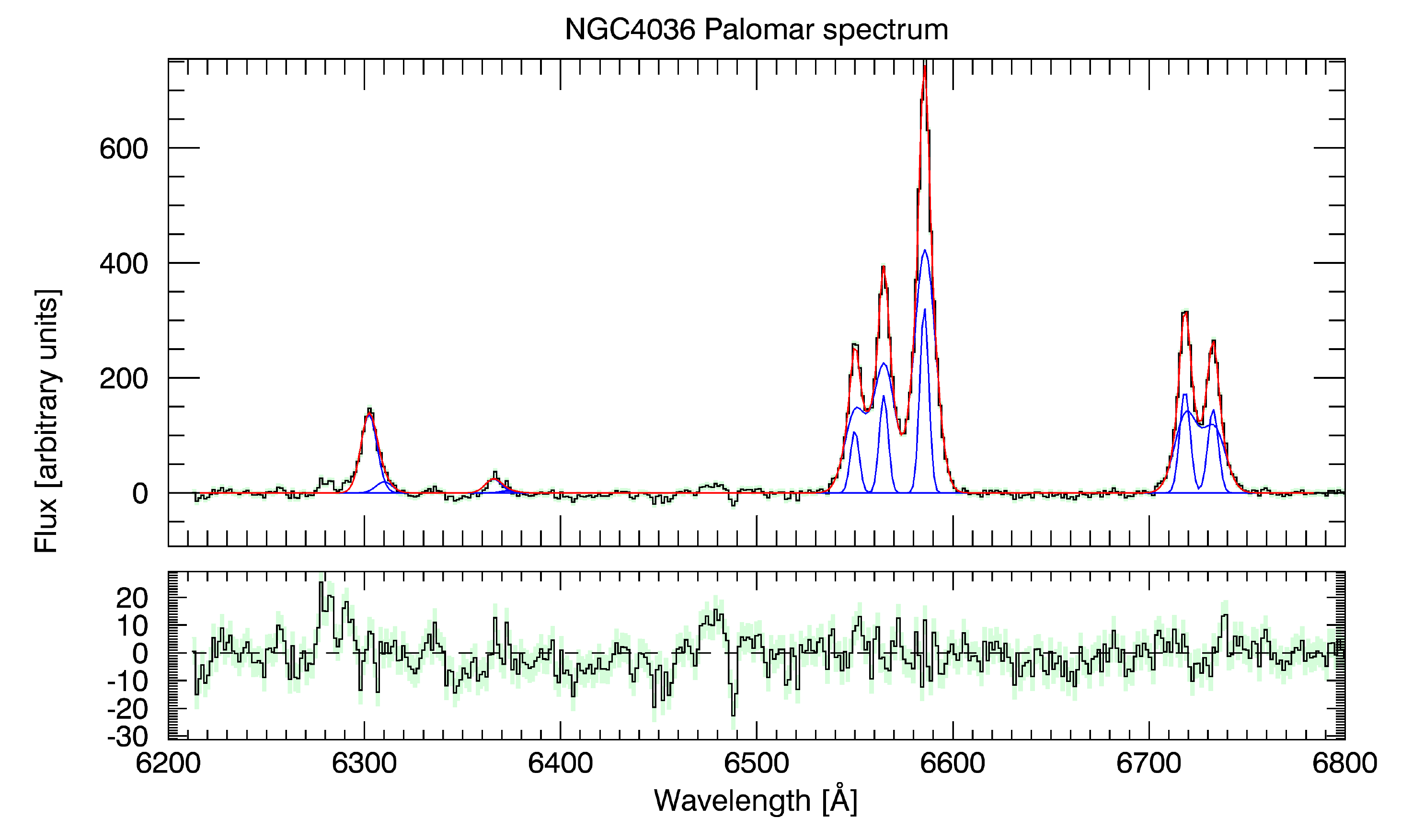}
  \includegraphics[width=9cm,angle=0]{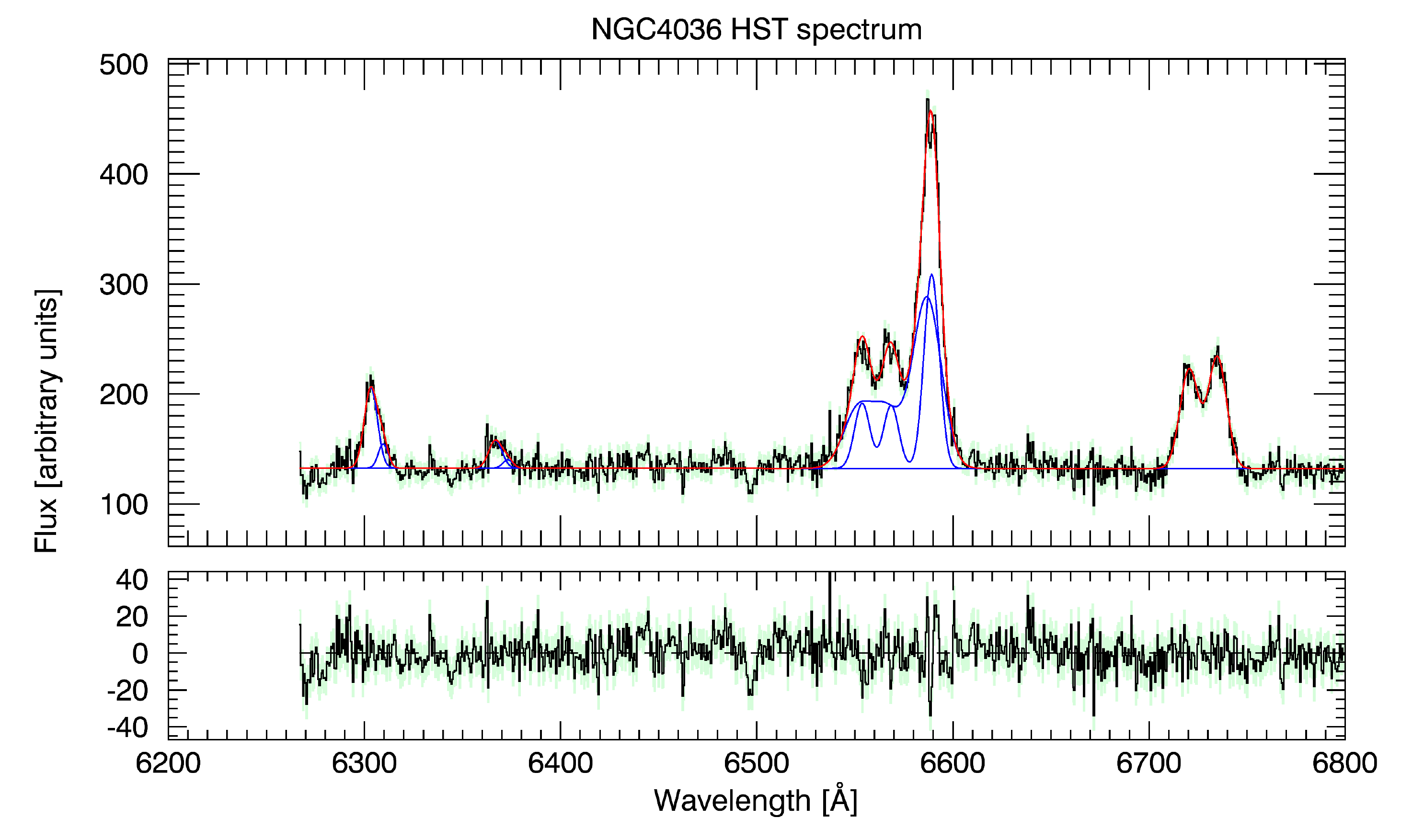}}
\caption{NGC~4036, as an example of the spectral analysis performed on a
  galaxies of the sample. We show the Palomar spectrum in the top panel
  (black), the best-fit starlight model (red) and, in the inset below, the
  residuals. In the middle panel, we present the fit to the emission lines (red lines) and the different components (blue lines). The HST is shown
  in the bottom panel with the same symbols used for the Palomar spectrum.}
\label{palomar}
\end{figure}

\section{Sample selection and analysis of the spectra}
\label{sample}

\begin{table*}
\caption{The sample galaxies and log of the HST/STIS observations.}
\begin{tabular}{l|c|c|c|c|c|l|l|l}
\hline
Name &  D & $\sigma_{\star}$ & log L$_{\oiii}$ & Slit width
& Obs. Date& Obs. Id & Exp. Time & Scale \\
     &  [Mpc] & [km s$^{-1}$] & [erg s$^{-1}$] & [arcsec] & & & [s] & [\AA/pix]\\
\hline
\multicolumn{9}{l}{Liners Type 1}\\
\hline
NGC~3031  &  1.4  & 162 &  37.72  &0.1$\arcsec$ &1999-07-14 & O51301010        &1000 \,\,\,\,\,\,\,\,  & 0.55 \,\,\,\,\,\,\,\,\\
NGC~3998  & 21.6  & 305 &  39.62  &0.1$\arcsec$ &2002-04-07 & O6N902010 to 40  & 520 \,\,\,\,\,\,\,\,  & 1.11 \,\,\,\,\,\,\,\,\\
NGC~4203  &  9.7  & 167 &  38.53  &0.2$\arcsec$ &1999-04-18 & O4E010010 to 30  &2779 \,\,\,\,\,\,\,\,  & 0.55 \,\,\,\,\,\,\,\,\\
NGC~4450  & 16.8  & 135 &  38.78  &0.2$\arcsec$ &1999-01-31 & O4E016010 to 30  &2697 \,\,\,\,\,\,\,\,  & 0.55 \,\,\,\,\,\,\,\,\\
NGC~4579  & 16.8  & 165 &  39.42  &0.2$\arcsec$ &1999-04-21 & O57208040	       &2692 \,\,\,\,\,\,\,\,  & 0.55 \,\,\,\,\,\,\,\,\\ 
\hline                              
\multicolumn{9}{l}{Liners Type 2}\\
\hline                              
NGC~0315  & 65.8  & 304 &  39.44  &0.1$\arcsec$ &2000-06-18 & O5EE02060 to 70 &  500 \,\,\,\,\,\,\,\,  & 1.11 \,\,\,\,\,\,\,\,\\
NGC~1052  & 17.8  & 215 &  40.10  &0.2$\arcsec$ &1999-01-02 & O57203050       & 1974 \,\,\,\,\,\,\,\,  & 0.55 \,\,\,\,\,\,\,\,\\
NGC~1961  & 53.1  & 165 &  39.11  &0.2$\arcsec$ &2001-12-27 & O6F450030 to 40 & 4963 \,\,\,\,\,\,\,\,  & 1.11 \,\,\,\,\,\,\,\,\\ 
NGC~2787  & 13.0  & 202 &  38.37  &0.2$\arcsec$ &1998-12-05 & O4E002010 to 30 & 2012 \,\,\,\,\,\,\,\,  & 0.55 \,\,\,\,\,\,\,\,\\
NGC~2911  & 42.2  & 243 &  39.34  &0.1$\arcsec$ &1998-12-03 & O4D311020       &  486 \,\,\,\,\,\,\,\,  & 1.11 \,\,\,\,\,\,\,\,\\ 
NGC~3642  & 27.5  &  85 &  38.96  &0.2$\arcsec$ &2000-10-13 & O5H720030 to 40 &  864 \,\,\,\,\,\,\,\,  & 0.55 \,\,\,\,\,\,\,\,\\
NGC~4036  & 24.6  & 215 &  39.16  &0.2$\arcsec$ &1999-03-25 & O57206030       & 2896 \,\,\,\,\,\,\,\,  & 0.55 \,\,\,\,\,\,\,\,\\
NGC~4143  & 17.0  & 205 &  38.81  &0.2$\arcsec$ &1999-03-20 & O4E009010 to 30 & 2856 \,\,\,\,\,\,\,\,  & 0.55 \,\,\,\,\,\,\,\,\\
NGC~4258  &  6.8  & 148 &  38.76  &0.2$\arcsec$ &2001-03-16 & O67104030	      & 1440 \,\,\,\,\,\,\,\,  & 0.55 \,\,\,\,\,\,\,\,\\ 
NGC~4278  &  9.7  & 261 &  38.88  &0.1$\arcsec$ &2000-05-11 & O57207030       & 3128 \,\,\,\,\,\,\,\,  & 0.55 \,\,\,\,\,\,\,\,\\
NGC~4477  & 16.8  & 177 &  38.82  &0.2$\arcsec$ &1999-04-23 & O4E018010 to 30 & 2613 \,\,\,\,\,\,\,\,  & 0.55 \,\,\,\,\,\,\,\,\\
NGC~4486  & 16.8  & 332 &  39.07  &0.1$\arcsec$ &2011-06-07 & OBIm01030 to 50 & 1717 \,\,\,\,\,\,\,\,  & 0.55 \,\,\,\,\,\,\,\,\\ 
NGC~4548  & 16.8  & 113 &  38.11  &0.2$\arcsec$ &1999-04-26 & O4E020010 to 30 & 2673 \,\,\,\,\,\,\,\,  & 0.55 \,\,\,\,\,\,\,\,\\ 
NGC~5005  & 21.3  & 172 &  39.41  &0.2$\arcsec$ &2000-12-24 & O5H741030       &  734 \,\,\,\,\,\,\,\,  & 0.55 \,\,\,\,\,\,\,\,\\
NGC~5077  & 40.6  & 255 &  39.52  &0.1$\arcsec$ &1998-03-12 & O4D305020       &  418 \,\,\,\,\,\,\,\,  & 1.11 \,\,\,\,\,\,\,\,\\ 
NGC~5377  & 31.0  & 170 &  38.81  &0.1$\arcsec$ &2011-03-04 & OBIB05010 to 60 & 4510 \,\,\,\,\,\,\,\,  & 0.55 \,\,\,\,\,\,\,\,\\ 
NGC~6500  & 39.7  & 214 &  39.90  &0.1$\arcsec$ &1998-11-03 & O4D307030       &  236 \,\,\,\,\,\,\,\,  & 1.11 \,\,\,\,\,\,\,\,\\ 
\hline                              
\multicolumn{9}{l}{Seyferts Type 1}\\
\hline                              
NGC~3227  & 20.6  & 136   & 40.68 &0.2$\arcsec$ &1999-01-31 &O57204040	       &1890 \,\,\,\,\,\,\,\,  & 0.55 \,\,\,\,\,\,\,\,\\ 
NGC~3516  & 38.9  & 181   & 40.80 &0.2$\arcsec$ &2000-06-18 &O56C01020 to 40   &2116 \,\,\,\,\,\,\,\,  & 0.55 \,\,\,\,\,\,\,\,\\
NGC~4051  & 17.0  &  89   & 40.18 &0.2$\arcsec$ &2000-03-12 &O5H730030 to 40   & 864 \,\,\,\,\,\,\,\,  & 1.11 \,\,\,\,\,\,\,\,\\ 
\hline                              
\multicolumn{9}{l}{Seyferts Type 2}\\
\hline                              
NGC~1358  & 53.6  & 222   & 40.84 &0.2$\arcsec$ &2002-01-25 &O6BU03010 to 30   &3000 \,\,\,\,\,\,\,\,  & 0.55 \,\,\,\,\,\,\,\, \\ 
NGC~1667  & 61.2  & 169   & 40.79 &0.2$\arcsec$ &2001-10-14 &O6BU04010 to 30   &3000 \,\,\,\,\,\,\,\,  & 0.55 \,\,\,\,\,\,\,\, \\
NGC~2273  & 28.4  & 149   & 40.43 &0.2$\arcsec$ &2001-11-04 &O6BU05010 to 30   &3266 \,\,\,\,\,\,\,\,  & 0.55 \,\,\,\,\,\,\,\, \\ 
NGC~3982  & 17.0  &  73   & 39.83 &0.2$\arcsec$ &1998-04-11 &O4E006010 to 30   &2997 \,\,\,\,\,\,\,\,  & 0.55 \,\,\,\,\,\,\,\,\\  
NGC~4501  & 16.8  & 167   & 39.10 &0.2$\arcsec$ &1999-04-26 &O4E019010 to 30   &2673 \,\,\,\,\,\,\,\,  & 0.55 \,\,\,\,\,\,\,\, \\  
NGC~4698  & 16.8  & 149   & 38.81 &0.2$\arcsec$ &1999-04-26 &O4E022010 to 30   &2673 \,\,\,\,\,\,\,\,  & 0.55 \,\,\,\,\,\,\,\, \\  
NGC~5194  &  7.7  &  96   & 38.91 &0.2$\arcsec$ &2002-04-03 &O6FM01010 to 50   &10467 \,\,\,\,\,\,\,\,  & 1.11 \,\,\,\,\,\,\,\,\\
NGC~6951  & 24.1  & 128   & 38.69 &0.2$\arcsec$ &2000-06-17 &O5H752030 to 40   &864 \,\,\,\,\,\,\,\,  & 0.55 \,\,\,\,\,\,\,\, \\ 
\hline
\end{tabular}
\label{table0}
\medskip

\small{(1) Object name;
  (2) distance of the source from \citet{ho95};  (3) stellar
  velocity dispersion from \citet{ho09a};  
  (4) Logarithm of the nuclear [OIII] emission line luminosity from  \citet{ho97}; from (5) to (9) HST data description.}
\end{table*}
\medskip
  
\citet{filippenko85} and \citet{ho95} present the results of an optical
spectroscopic survey, performed with the Palomar 5 m Hale telescope, of 486
nearby bright galaxies ($B_T \le$ 12.5 mag) located in the northern sky
(hereafter referred to as the Palomar survey). We consider the 106 galaxies in
this survey robustly classified in \citet{ho97} as Seyfert, LINER, or
transition objects between these two classes (according to the spectroscopic
criteria proposed by \citealt{veilleux87}). We searched the Hubble Legacy
Archive (HLA) for spectroscopic observations performed by HST/STIS with G750M
grating\footnote{The spectral resolution of the G750L grism is insufficient
  for our purposes.} and centered to cover the \Ha\ emission line. We found 42
observations, but we discarded nine sources (namely NGC~3254, NGC~4261,
NGC~4314, NGC~4594) because of the poor quality of the spectra, NGC~4138 and
NGC~4374 because the slit was not centered on the nucleus, and NGC~3368,
NGC~4636, and NGC~4736 because they are likely not AGN (see
\citealt{balmaverde15} for details). Our final sample is therefore composed of
33 sources, which we list in Table \ref{table0}.

All the reduced and calibrated spectra of the Palomar survey are available
through the NASA Extragalactic Database (NED). The long slit spectra obtained
with the red camera cover the wavelengths between 6210 and 6860 \AA\, with a
resolution of $\sim$ 2.5 \AA. We use the Gandalf software \citep{sarzi06} to
model and subtract  starlight emission. The residuals contain the emission
lines, and the most prominent in the spectral range considered are
\oi$\lambda\lambda$6300,6363, \nii $\lambda\lambda$6548,6584, \Ha, and [S
  II]$\lambda\lambda$6716,6731. We fit these lines with the IDL routine {\sl
  mpfit} that employs a $\chi^2$ minimization procedure. We model each line
with two Gaussians to reproduce both the narrow core component and
the broad wings, and we derive the line properties by summing the two
components. The \nii\ doublet ratio is fixed to the expected value of 1:3
\citep{humphrey08}, while the widths of \nii\ and \Ha\ are assumed to be
equal. For the type 1 AGN we add a broad \Ha\ component with a broken
power-law profile, which can reproduce both asymmetric and skewed profiles
\citep{nagao06}.

We consider public HST/STIS spectra in the HLA (see Table \ref{table0}). The
G750M grism covers the range between 5450 and 10140 \AA\ with a resolving
power of R$\sim$ 5000, with 572 \AA\ covered in each tilt position. When
possible, we combine multiple observations to remove cosmic rays and bad
pixels. From the fully calibrated data, we extract the nuclear spectrum from a
synthetic aperture of 0\farcs15. In the HST spectra the continuum emission
does not show a significant presence of absorption features from stars. We
then simply reproduce the continuum in the HST spectra as a constant (or at
most with a first degree polynomial) without performing a starlight
subtraction. The fit to the emission lines follows the same strategy described
above.

For two of the type I Seyfert (namely NGC~3516 and NGC~4051), the BLR profile
is too complex to be reproduced with any analytical form and the measurements
of the narrow lines in the \nii+\Ha\ complex is compromised. For these objects, we adopted a different strategy: we only considered  the \oi\ and the \sii\ lines
by locally fitting  a second order polynomial to reproduce the continuum and
BLR wings.

The analysis of the HST spectra revealed the possible presence of another
LINER with a broad \Ha\ line, namely NGC~3642. This LINER shows an increase in
the narrow line widths from the Palomar to the HST data by a factor of more
than $\sim 3$, which is significantly larger than observed in the other
galaxies. We believe that this is an indication of the presence of a BLR and,
thus, we include this component in the model fitting.
 
As an example, we show in Fig. \ref{palomar} the Palomar (before and after
starlight subtraction) and the HST spectra of NGC~4036.  In Table
\ref{bigtable} we report fluxes and full widths at half maximum (FWHM) of each
line (measured on the sum of the two Gaussian components).

\section{Emission lines properties}
\label{gbcfr}

\begin{figure}
\centering{
\includegraphics[scale=0.45,angle=0]{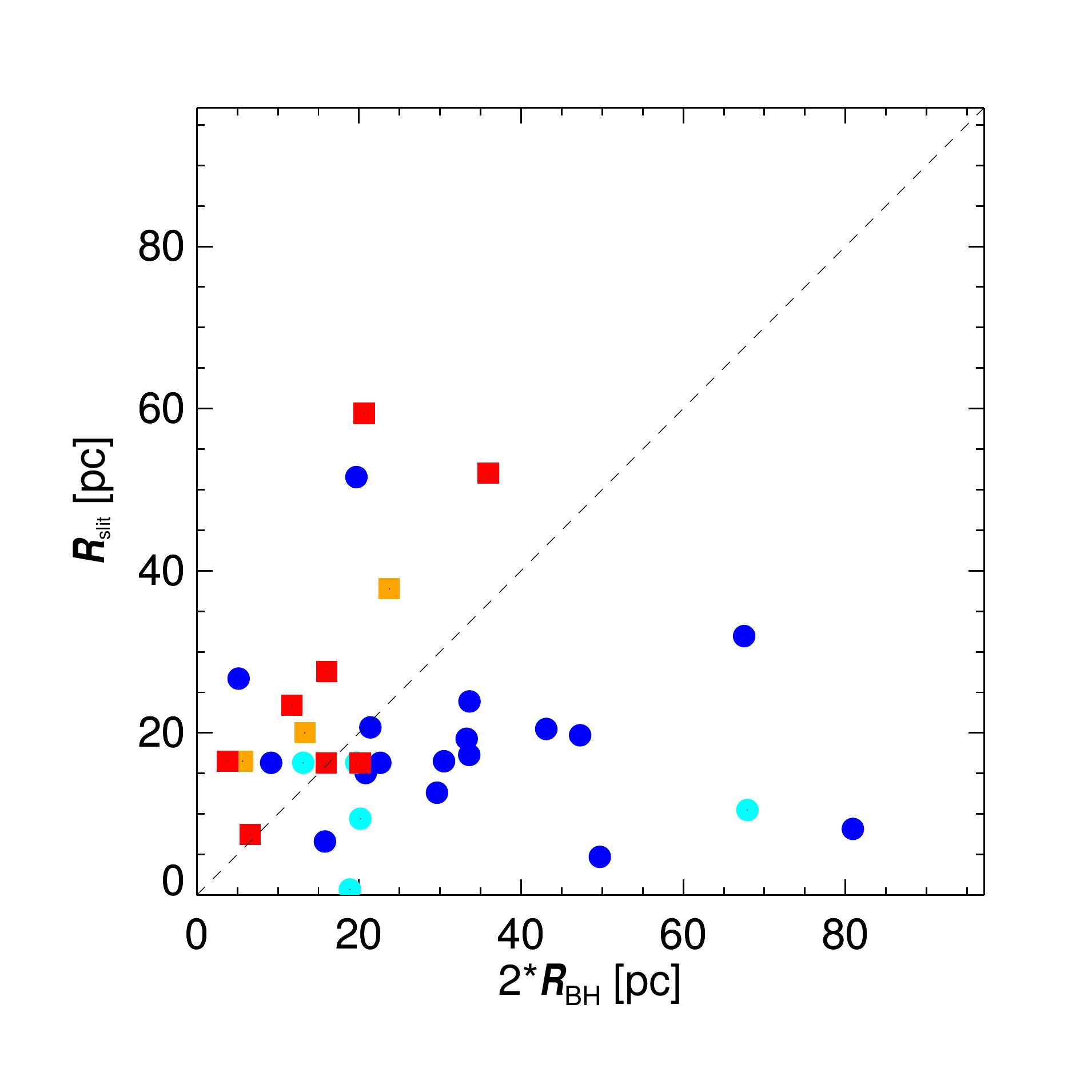}}
\caption{Projected size of the region covered by the HST slit versus the
  diameter of the black hole sphere of influence (both in parsec). Red squares
  represent Seyferts2, orange squares Seyferts1, blue dots LINERs2, and cyan
  dots LINERs1.}
\label{sphere}
\end{figure}

The highest spatial resolution of HST offers the unique opportunity to
explore the properties of the innermost region of the NLR (hereafter
INLR). Indeed, the HST aperture is 0.2\arcsec (or even 0.1\arcsec) $\times$
0.15\arcsec, which is much smaller than the 2\arcsec $\times$ 4\arcsec aperture of the
Palomar survey, and provides us with a much better spatial resolution.  As a
result, the flux of the \nii\ line in HST spectra is lower than in the Palomar
spectra: the ratio between these two fluxes ranges from 1 to 20 in LINERs and
from 4 to 40 in Seyferts.

In Fig. \ref{sphere}, we compare the linear dimension of the region covered by
the HST slit with the diameter of the black hole (BH) sphere of influence,
whose radius is defined as $R_{\rm BH}=GM_{\rm BH}/\sigma_\star^2$, where
$\sigma_\star$ is the stellar velocity dispersion. This is the region within
which the gravitational potential is dominated by the presence of the
supermassive BH (whose mass is estimated from the values of $\sigma_\star$
reported in Table \ref{table0}, adopting the \citealt{tremaine02}
relation). In the LINERs, the diameter of the BH sphere of influence is between
4 and 81 pc with a median of 21 pc, larger than the median distance covered by
the slit (16 pc). Indeed, in all but three LINERs, the HST spectra probe a
region confined within the BH sphere of influence.

This is not the case for the Seyferts, in which   the sphere of
influence is only comparable to the slit size in three cases. In general, this region is
unresolved, mainly because of the lower BH masses in these objects with respect to
LINERs.

\begin{figure*}
\centering{
\includegraphics[scale=0.45,angle=0]{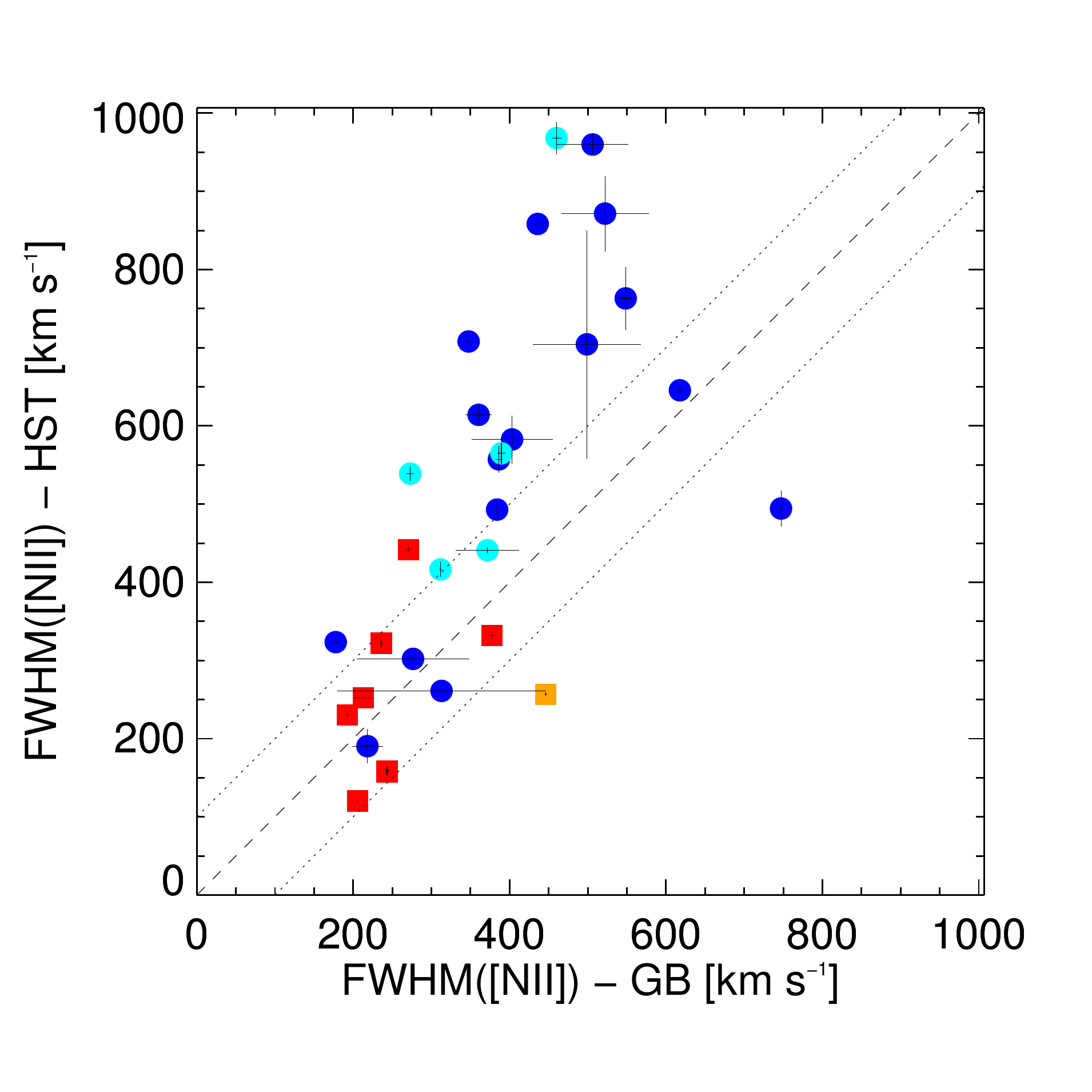}
\includegraphics[scale=0.45,,angle=0]{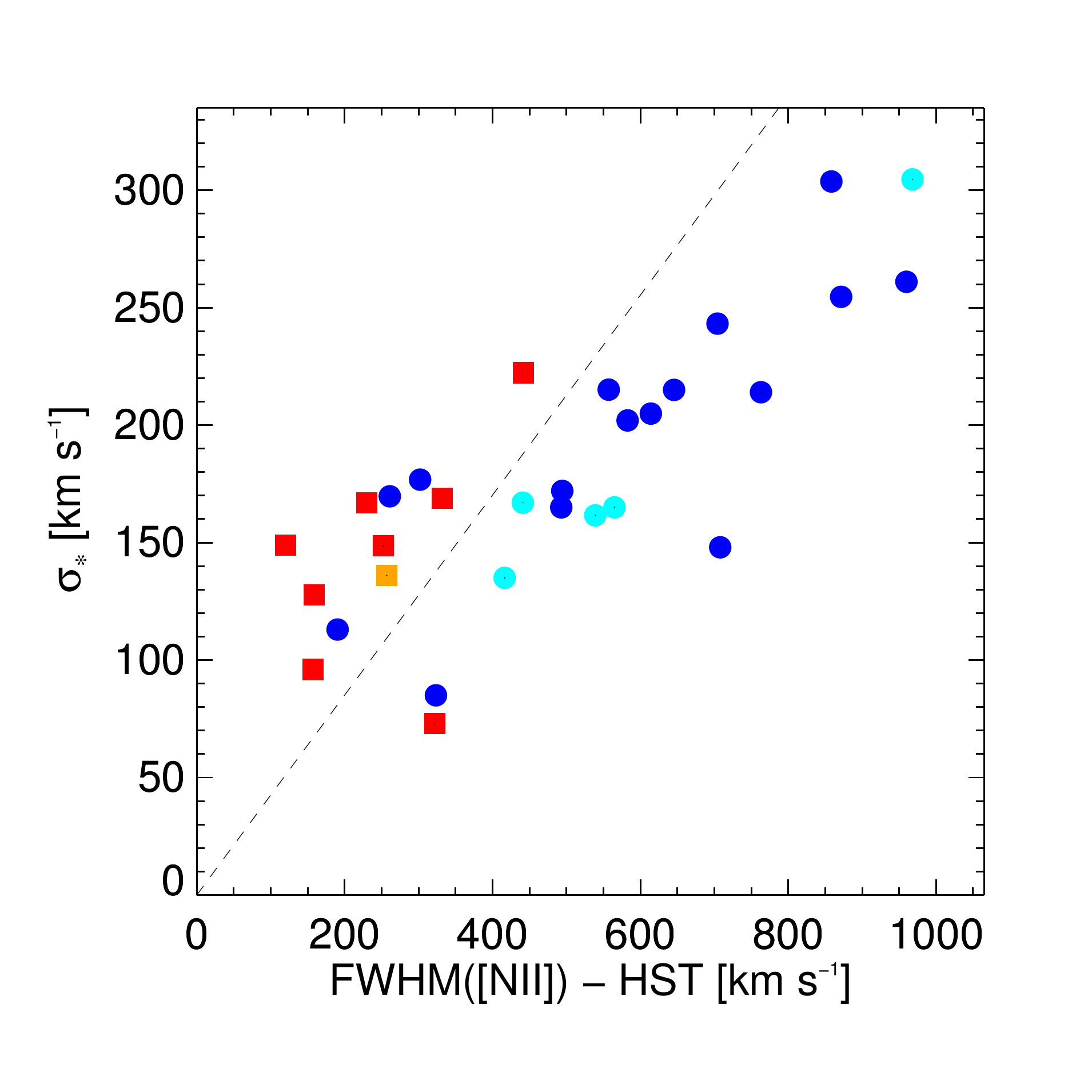}}
\caption{Left panel: comparison of the FWHM of \nii\ emission line from HST
  and Palomar spectra. The dashed line is the bisectrix of the plane and the
  dotted lines enclose a region of $\pm$100 \kms centered on the
  bisectrix. Symbols as in Fig. \ref{sphere}.  NGC~4486 falls outside
  the plotting range.  Right panel: stellar velocity dispersion in \kms versus
  the FWHM of the \nii\ emission line. The dashed line corresponds to FWHM\nii
  = 2.35 $\times$ $\sigma_\star$.}
\label{n2n2g}
\end{figure*}

\subsection{Ground-based vs. HST measurements}

Initially we focus on the \nii\ emission line (the brightest line in these
spectra) as seen in the HST and ground-based spectra. In Fig. \ref{n2n2g}, left
panel, we compare the FWHM\nii\ measurements in the Palomar and HST
spectra. We do not see any significant change of the FWHM moving from 
ground-based to  HST spectra in the Seyferts (the median are 243 and 252
\kms, respectively; see Table \ref{tab1}).  Instead, in the LINERs of our
sample, the median FWHM\nii\ grows from 389 to 564 \kms. This is likely because in LINERs, opposite from what happens in Seyferts, the HST slit is
well within the sphere of influence of the BH, where the velocities sharply
increase toward the center.

The prevalence of gravitational motions in the NLR of the objects considered
(as opposed to, e.g., a kinematic dominated by the effects of the AGN) is
supported by the comparison between the central stellar velocity dispersion
and FWHM\nii\ measured from HST spectra. Considering only the LINERs, the
FWHM\nii\ correlates with the stellar velocity dispersion (with a linear
correlation coefficient of 0.8 and a slope of $\sim 3$); see Fig. \ref{n2n2g},
right panel. The line widths are in general higher than $2.35 \times
\sigma_\star$ (where 2.35 is the conversion factor from dispersion to FWHM) by
just $\sim$ 25\%. Also, in the Seyferts FWHM\nii\ correlates with
$\sigma_\star$, but with a shallower slope.

In the assumption of the dominance of gravitational motions and, for the
LINERs, the prevalence of the BH potential, we can convert the observed
velocities into distances by assuming Keplerian rotation, i.e., $R_{\rm
  INLR}\sim GM_{\rm BH}/(f\, {\rm FWHM}^2)$.\ The parameter  $f$ is a geometric factor that
depends on the structure of these innermost regions of the NLR. We adopt $f
=\sqrt{3}/2.35$ as valid for an isotropic distribution of randomly moving clouds
(\citealt{peterson04})\footnote{\citet{peterson04} adopted $f=\sqrt{3}$ but
  they measured the line dispersion instead of the FWHM. This value is
  equivalent to $f =\sqrt{3}/2$ in \citet{netzer90} that used FWHM
  velocities. An empirical derived value of $f$=5.5 is reported by
  \citealt{onken04}.}.  We find a range of radii between 1$< R_{\rm FW[NII]}<$
10 parsec with a median radius of 3.0 parsec.

\begin{figure}
\includegraphics[scale=0.45,angle=0]{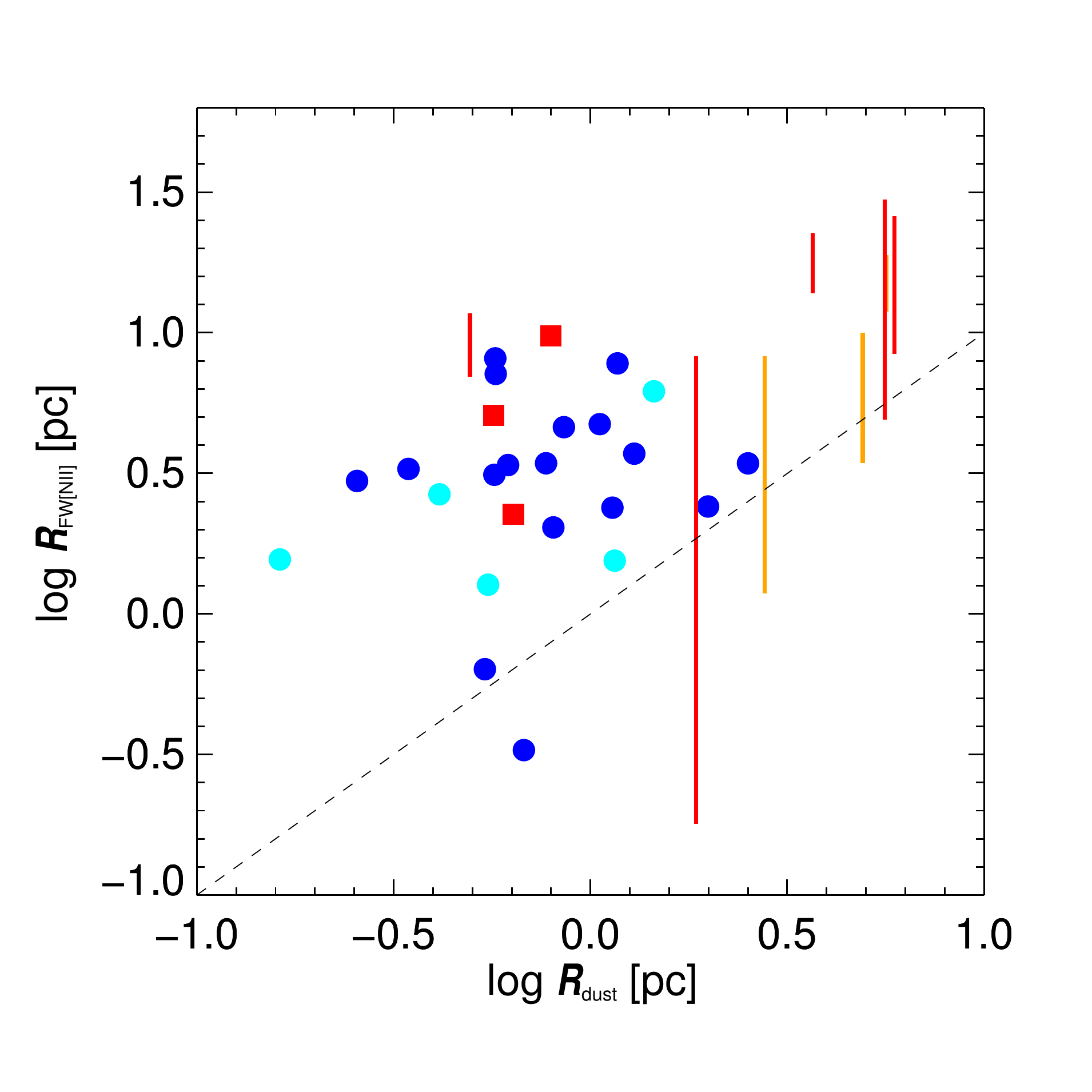}
\caption{Logarithm of the dust sublimation radius versus size of the INLR for the LINERs and
  for the three Seyfert with resolved spheres of influence, all in pc units. 
  For the remaining Seyferts we draw  the allowed ranges of the INLR size as vertical segments (see text for details).}
\label{torus}
\end{figure}

As a useful comparison, we calculate the dust sublimation radius $R_{\rm d}$
for the LINERs  using the relationship given by \citet{elitzur08}, i.e.,
$R_{\rm d}=12 \times \sqrt{L_{\rm Bol}/L_{45}}$ pc; the bolometric luminosity
is derived from the \oiii\ line luminosity as $L_{\rm Bol}=3500 \times
L_{[~OIII]}$ \citep{heckman04}. In Fig. \ref{torus}, we compare the radius of
the INLR (estimated for all LINERS and for the three Seyfert with resolved
spheres of influence) with the dust sublimation radius: these two regions are
essentially the same size in all objects, where $R_{\rm FWHM[NII]}$ is typically a factor four larger, although with a substantial spread.

In the other Seyferts, the sphere of influence is not resolved but we can
nonetheless derive constraints on the line emitting gas location. An upper
limit to the size of the INLR is simply related to the slit size, and a lower
limit is imposed by the radius at which the line width would exceed the
observed values: if the INLR was smaller that this limit we would observe
broader lines. The resulting ranges, shown as vertical segments in
Fig. \ref{torus}, are typically of $\sim$ 10 pc, generally slightly larger than
the dust sublimation radius.

The mass of ionized gas in the INLR can be
estimated from the following observed Balmer line luminosity \citep{osterbrock89}:

$$M_{\rm gas}={{m_p+0.1m_{\rm He}}\over{n_e\alpha_{\rm H\beta}^{eff}h\nu_{\rm H\beta}}} L_{\rm H\beta.}$$

We converted the measured $L_{\rm H\alpha}$ into $L_{\rm H\beta}$ by adopting
a standard 3.1 ratio (e.g.,\citealt{gaskell84}). By assuming a temperature of
10$^4$ K, i.e., a recombination coefficient $\alpha_{\rm
  H\beta}^{eff}=3.03~10^{-14}$ cm$^3$ s$^{-1}$, and a density of $n_{\rm e} =
10^4 {\rm cm}^{-3}$, we find a range $M_{\rm gas}$ between 10$^{2}$ and
10$^{5} \,\,(10^4/n{\rm _e}) M_\sun$ with a median value of 10$^{3.8}$
M$_\sun$.

\subsection{Comparison of line widths and intensities}

\begin{figure*}
\centering{ \includegraphics[scale=0.40,angle=0]{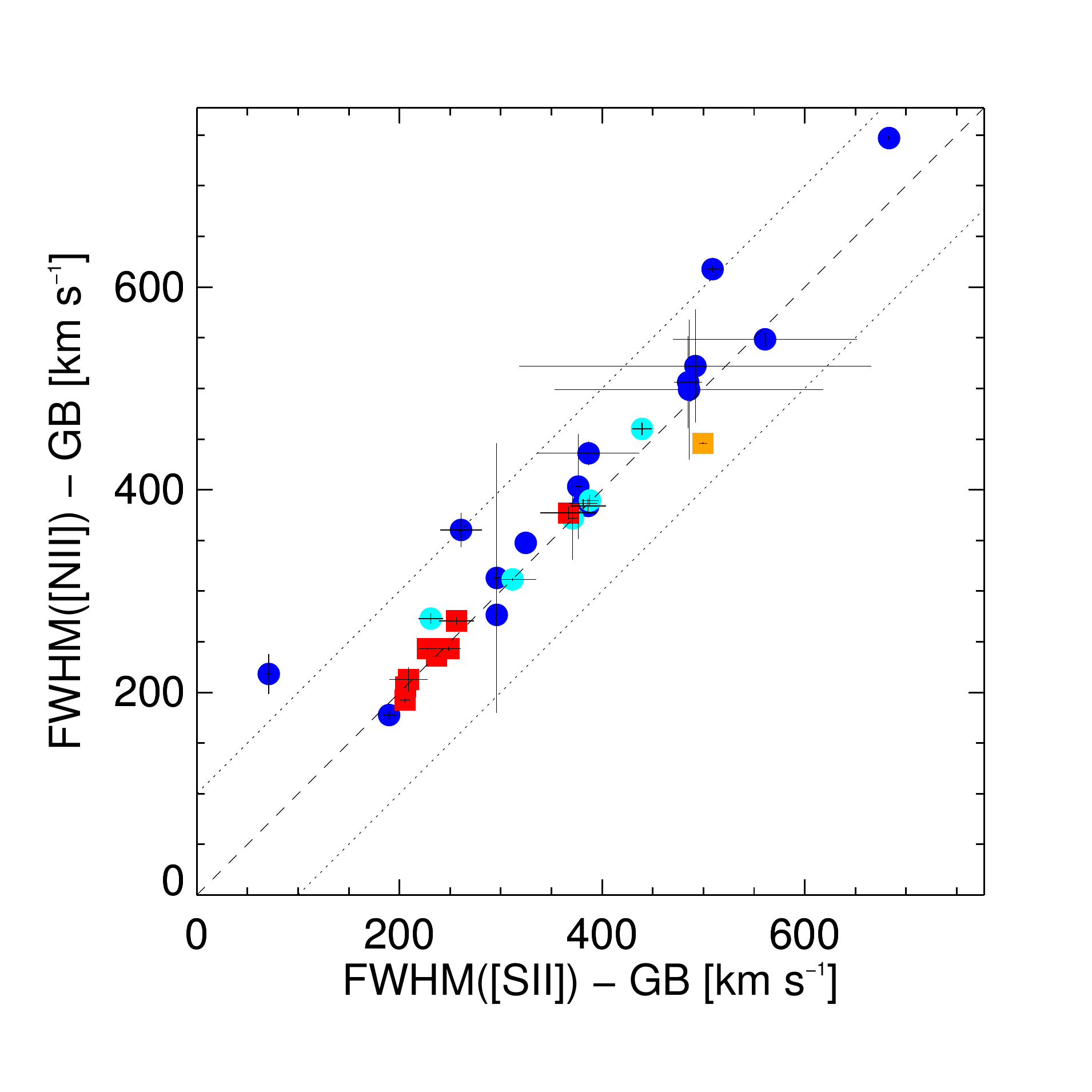}
  \includegraphics[scale=0.40,angle=0]{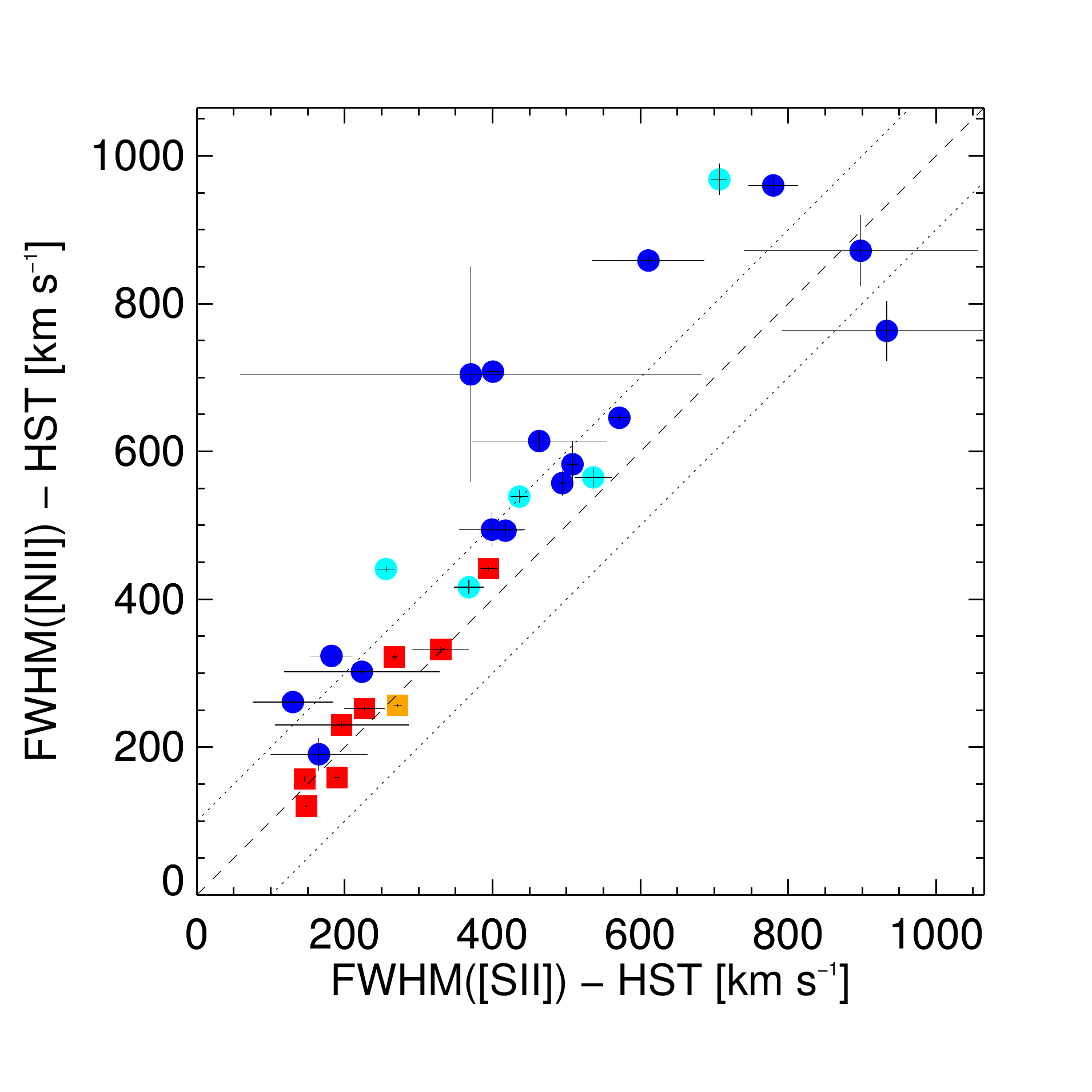}
 \includegraphics[scale=0.40,angle=0]{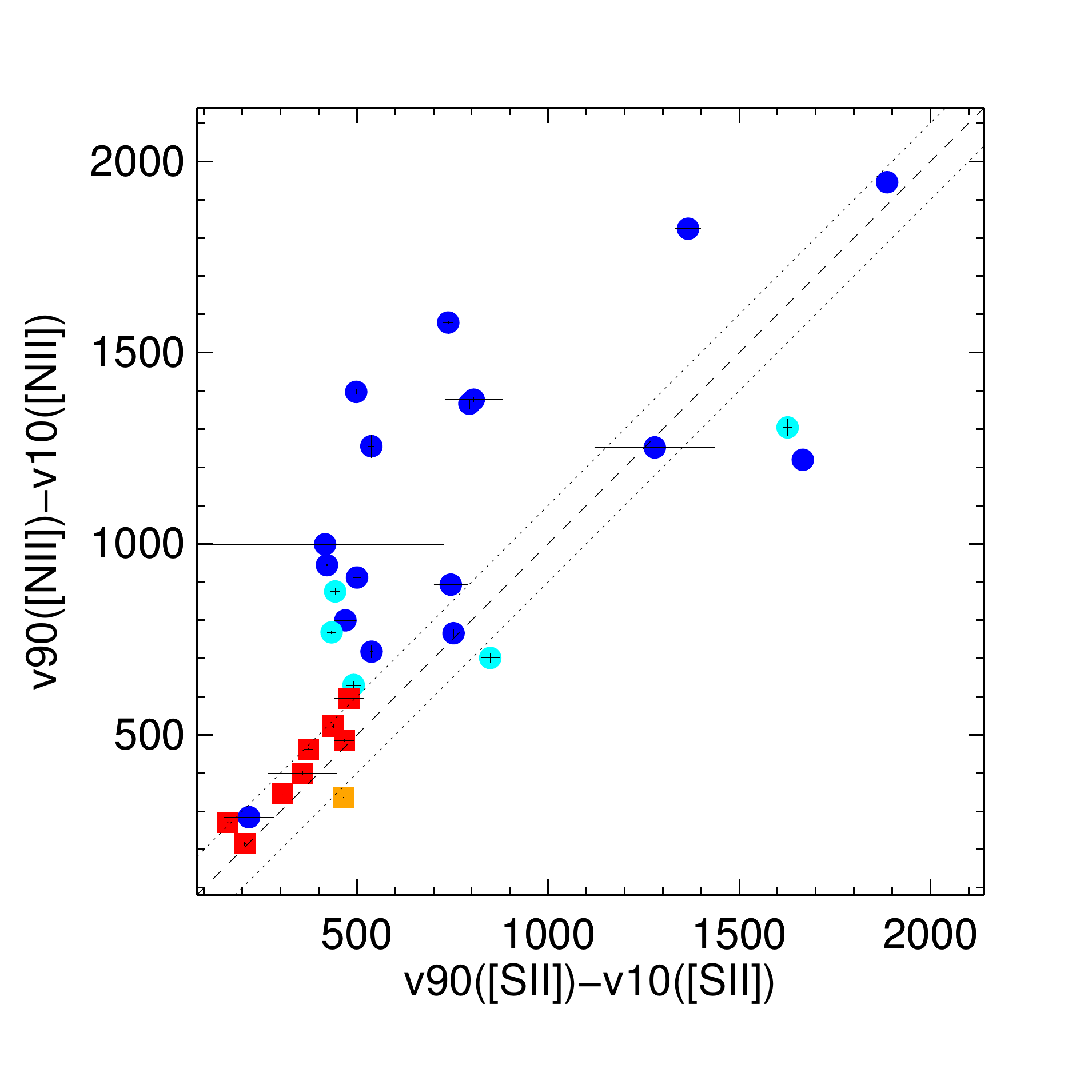}
  \includegraphics[scale=0.40,angle=0]{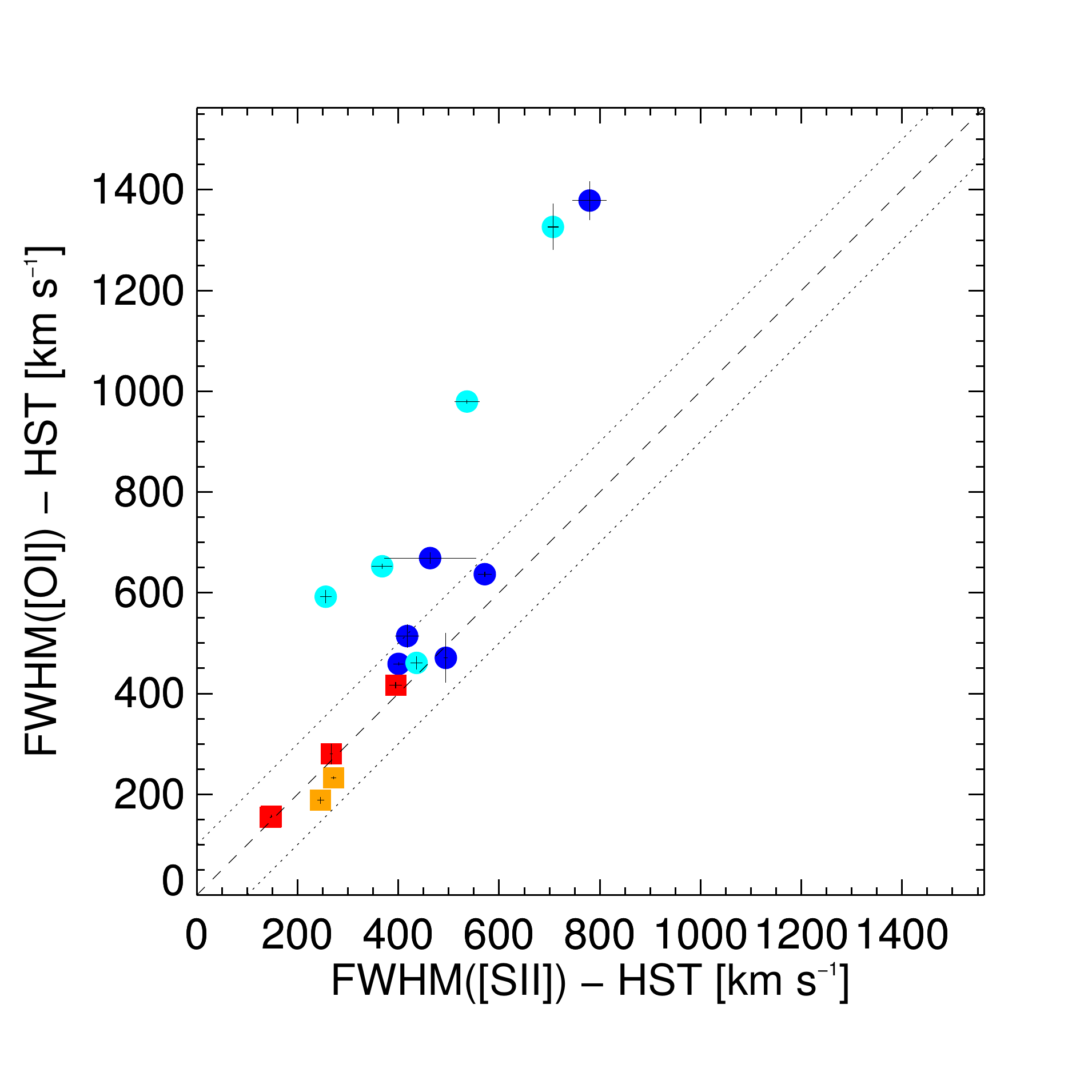}}
\caption{Comparison of the FWHM of the \nii\ and \sii\ emission lines from
  Palomar (top left) and HST (top right) spectra. NGC~4486 falls outside the
  plotting range.  In the bottom left panel, we  used the differences between
  the velocity defined at 10\% and 90\% of the total flux instead. In bottom right
  panel, we compare the FWHM of the \oi\ and \sii\ lines. Symbols as in
  Fig. 2.}
\label{n2s2}
\end{figure*}

We now analyze the behavior of the \sii and \oi emission lines and compare it
to the results obtained for \nii discussed above. In ground-based spectra the
\nii and \sii emission line do not show any remarkable differences in their
FWHM (see Table \ref{tab1}). Conversely, in the HST spectra the LINERs have a
median FWHM(\nii) 22\% larger than FWHM(\sii), 564 \kms against 463 \kms
(Fig. \ref{n2s2}, top panels). These differences are even more pronounced
considering the wings of the lines (bottom left panel), which are defined as the
difference in velocity between the 10th and 90th percentile (see
Sect. \ref{tomography} for a definition of the percentiles) with an increase
of 28\%. This broadening effect is limited to a change of $\sim$ 10\% in the
Seyferts.
  
The information about the \oi\ line is significantly less complete. In fact,
the \oi\ doublet, depending on the central wavelength selected for the HST
observations, might fall outside the spectra or, in other cases, only the
weakest \oi$\lambda$6363 line of the doublet is visible. Moreover, this line is
intrinsically fainter with respect to the other lines considered and the
quality of the spectra is often not sufficient to obtain a reliable fit to the
data. For these reasons, measurements of \oi\ are reported for only 18
objects. In Fig. \ref{n2s2}, bottom right panel, we plot FWHM(\oi) versus
FWHM(\sii). We find that for most of the LINERs the \oi\ velocity dispersion
is higher than \sii\ (the mean values in this subsample are 740 and 495 \kms,
respectively). Instead, for the Seyferts the FWHM of \oi\ and \sii\ do not
differ significantly.

\begin{figure}
\centering{
\includegraphics[scale=0.45,angle=0]{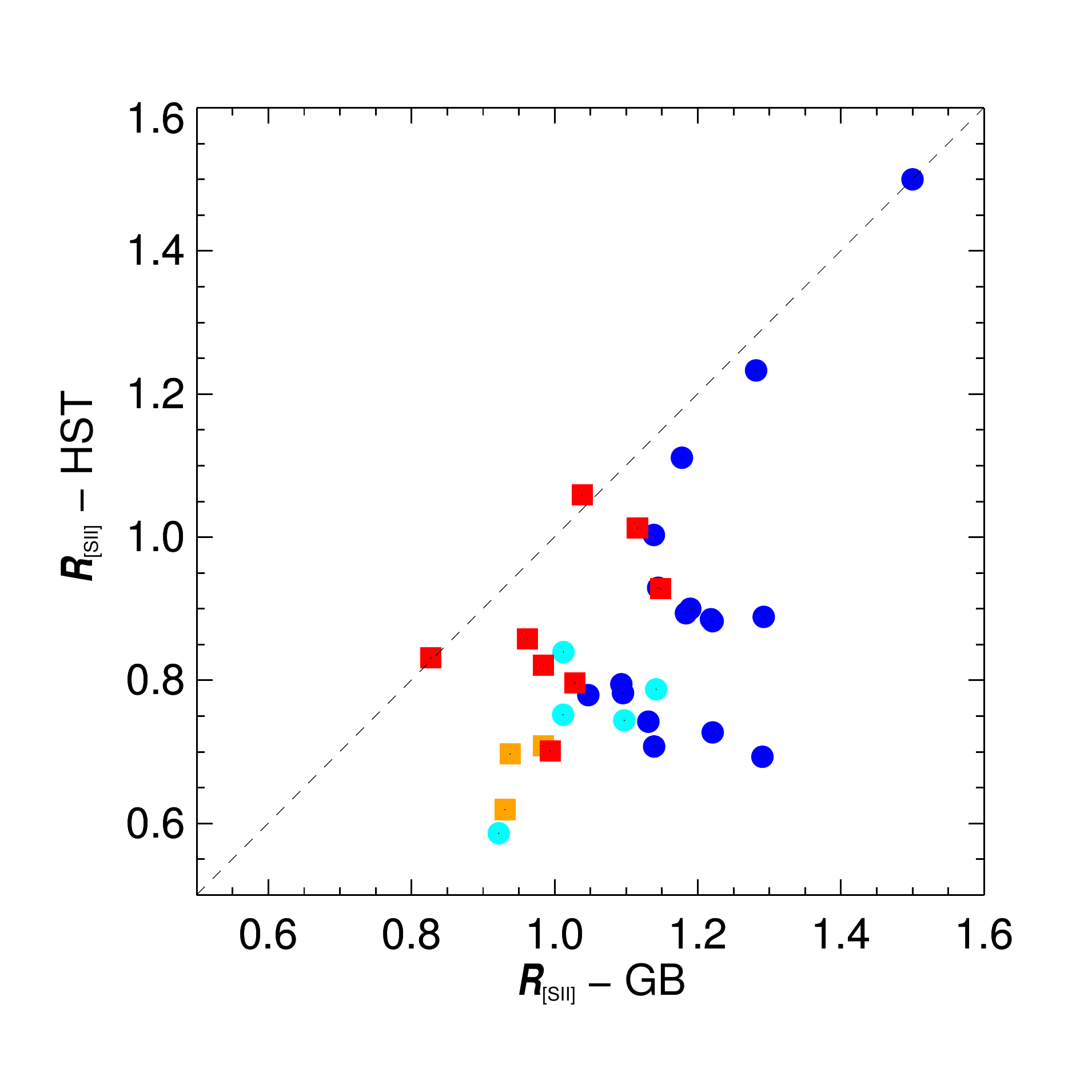}}
\caption{\sii$\lambda6716$/\sii$\lambda6731$ flux ratio measured from 
Palomar and HST data.}
\label{density}
\end{figure}

The intensity ratio of the two lines forming \sii\ doublet is commonly used to
derive the electron density of the gas.  This ratio decreases from
\sii$\lambda6716$/\sii$\lambda6731\sim$1.4 to $\sim$0.4 when the gas density
increases from $n_{\rm e}\sim$10 to $\sim 10^4$ cm$^{-3}$. For even higher
density the \sii\ doublet intensity ratio is insensitive to $n_{\rm e}$. In
Fig. \ref{density} we compare the \sii\ doublet ratios observed in HST and
Palomar spectra. In LINERs this ratio is always larger in the ground-based
than in HST spectra, with a median value increasing from $\sim$ 1.2 to
0.8. The corresponding $n_{\rm e}$ variation is from $\sim$ 10$^2$ to $\sim$
10$^3$ cm$^{-3}$. The value measured in HST spectra is often close to its high
density limit. For the Seyferts, there is a decrease in this ratio, but less
pronounced, from $\sim$1.0 to $\sim$ 0.8.

\begin{figure}
\includegraphics[scale=0.45,angle=0]{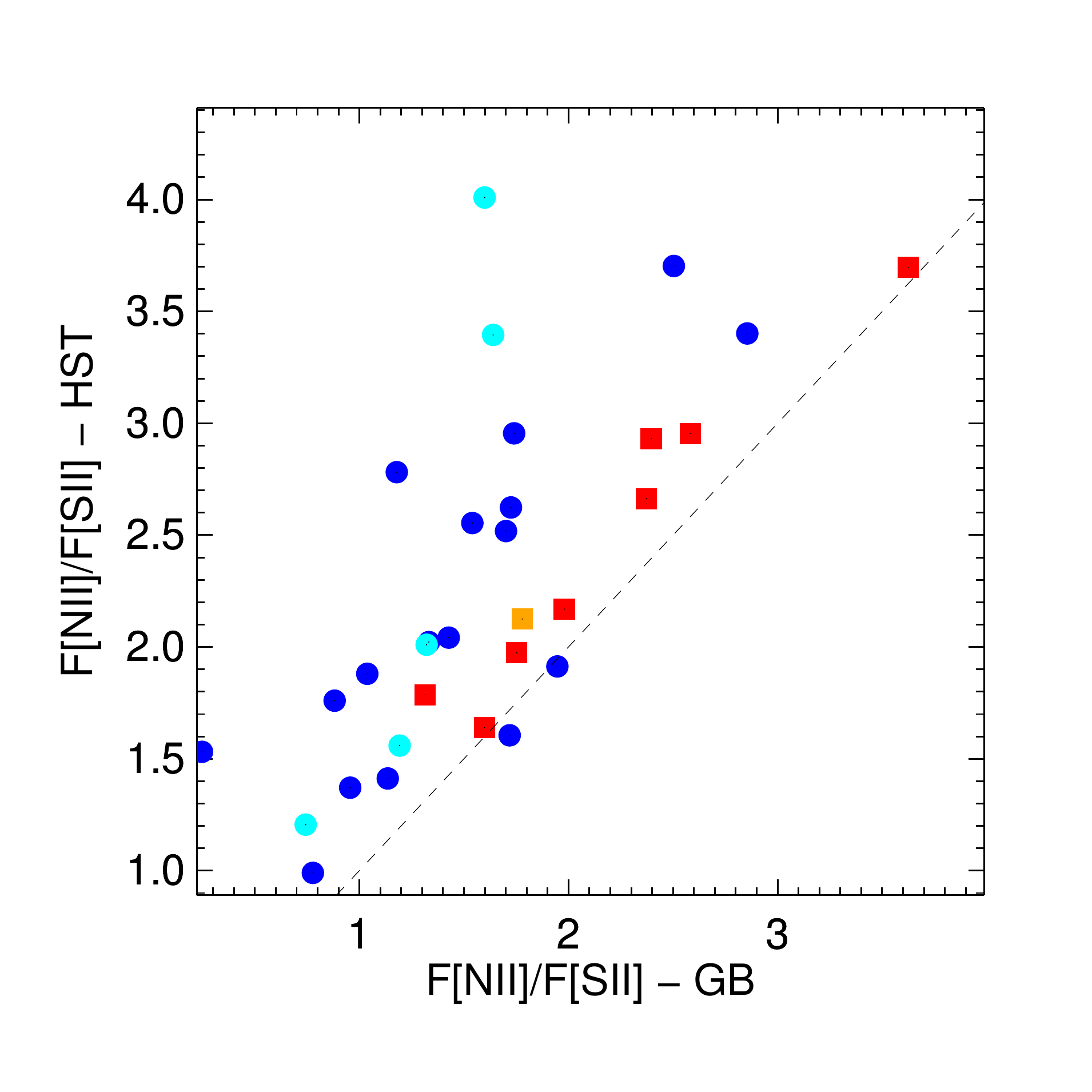}
\caption{Comparison of the ratios of the \nii\ and \sii\ fluxes measured from
  ground-based and HST spectra.}
\label{ratio1}
\end{figure}

We now consider the relative intensity of the \nii\ and \sii\ emission lines.
The ratios between the \nii and \sii\ fluxes in the Palomar spectra
$R_{\rm[N~II]/[S~II]}$ span a wide range of values, from $\sim$ 1 to $\sim$
4. In all but two LINERs there is a significant increase of
$R_{\rm[N~II]/[S~II]}$ from the Palomar data to the HST spectra, typically by
a factor two (see Fig. \ref{ratio1}).

\begin{table}
\caption{Line widths comparisons}
\begin{tabular}{l|c|c|c|c|}
\hline
     & \multicolumn{2}{|c|}{LINERs} & \multicolumn{2}{|c|}{Seyferts}\\ 
\hline
     & Median & Average & Median & average \\
\hline
\sii\ Palomar                    &    386 &    412  &    236 &    273 \\ 
\sii\ HST                          &    463 &    518  &    226 &    236 \\ 
\sii\ HST{\tiny(10-90)\%} &    738 &    794  &    358 &    345 \\ 
\nii\ Palomar                    &    389 &    440  &    243 &    270 \\ 
\nii\ HST                          &    564 &    583  &    252 &    252 \\ 
\nii\ HST{\tiny(10-90)\%} &    944  &   1041&    400 &    404 \\ 
\oi\  HST                          &    636 &    739  &    232 &    238 \\ 
\hline
\end{tabular}
\label{tab1}

\medskip
\small{All widhts are in \kms.} 
\end{table}
\medskip

\subsection{INLR tomography}
\label{tomography}
\begin{figure}
\centering{
\includegraphics[scale=0.4,angle=0]{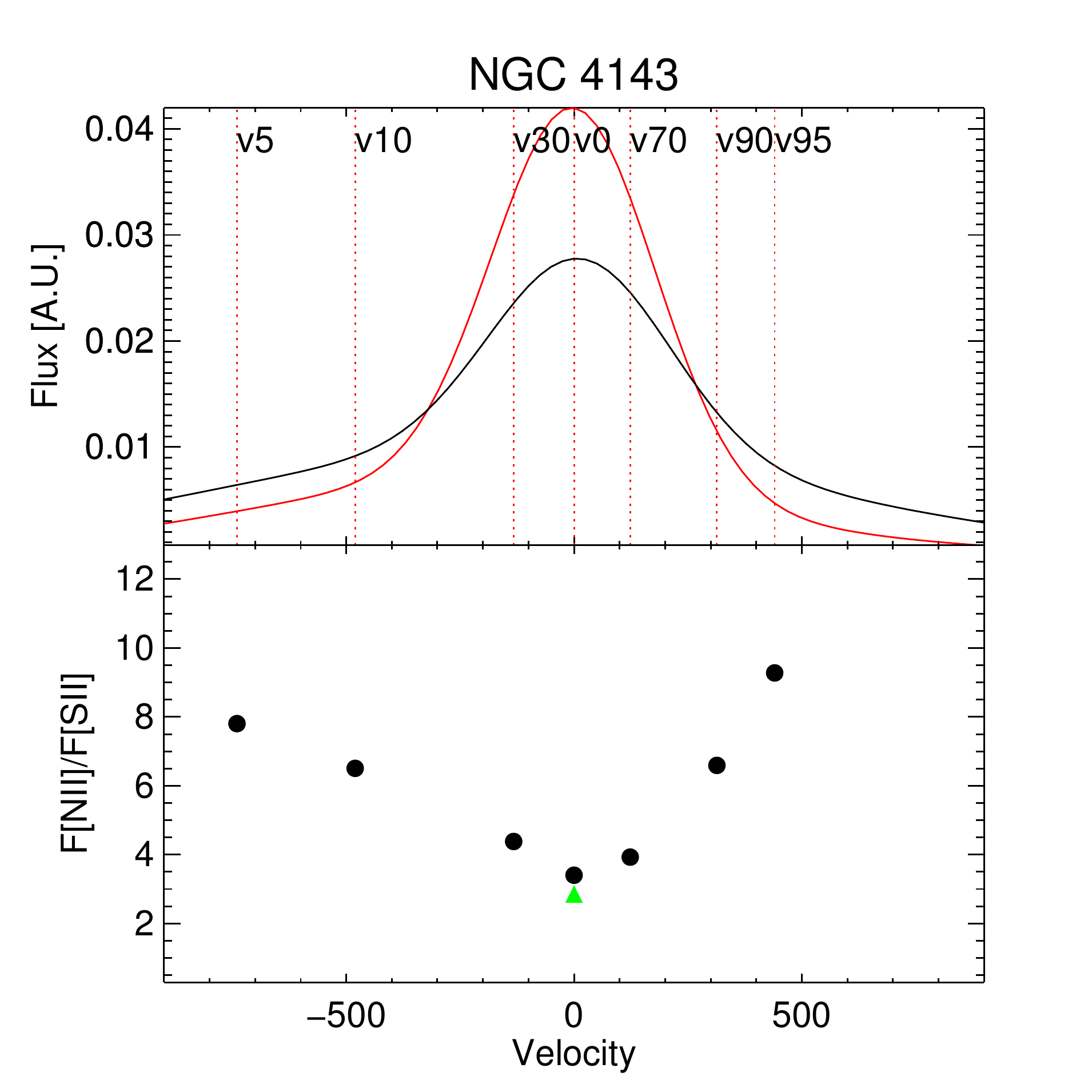}
\includegraphics[scale=0.4,angle=0]{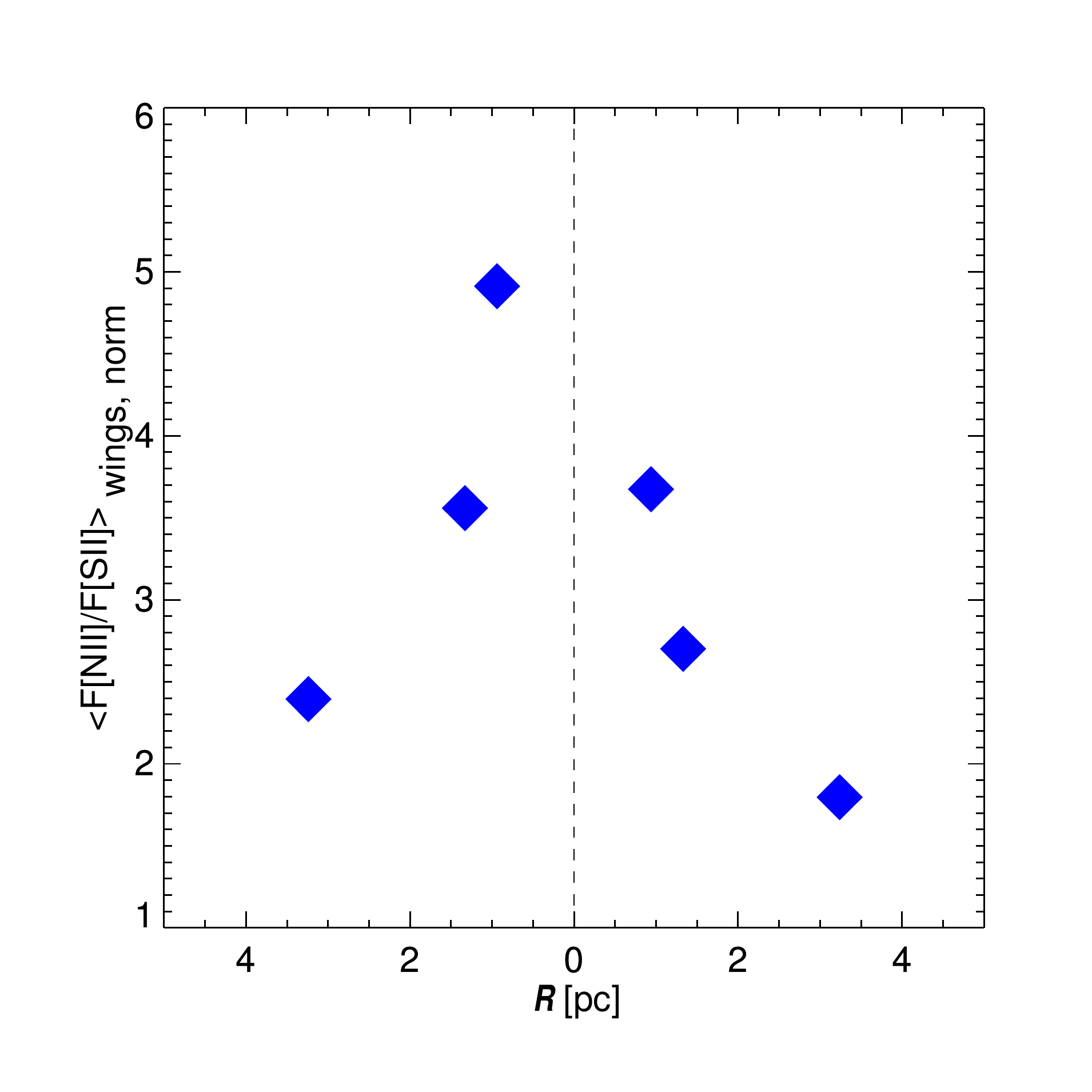}
}
\caption{\nii\ emission line (in black) superposed to the \sii\ emission line
  (in red) derived from the fit model for NGC~4143. We indicate the velocities at
  different percentiles of the total line flux (5\%, 10\% and 30\%, 70\%,
  90\%, and 95\%). The green triangle represent the ratio measured from the
  ground. In the bottom panel, we report the median of the
    $R_{\rm[N~II/[S~II]}$ ratio for all the LINERs in the wings, normalized to
      the values observed from the ground. This is compared with the median
      radius of the emitting region, based on the assumption of Keplerian
      rotation.}
\label{wings}
\end{figure}

So far we have considered mainly the global properties of the emission lines,
however, the lines are well resolved, particularly in the HST spectra.  This
enables us to consider separately the emission produced in different
wavelength ranges, i.e., at different velocities. In particular, the wings of
the lines sample regions located closer to the central black hole. In
Fig. \ref{wings}, upper panel, we show an example of this analysis for the
LINER NGC~4143. We superpose the \nii\ model line, obtained from the HST
spectrum, to the \sii model line. We align the peaks of the two lines and
normalize each line to its total flux. We consider velocities at different
percentile of the \sii flux. For example, $v_{5}$ is defined as the velocity
at which the line blue wing contains 5\% of the total flux. We similarly
define the velocities associated with the percentiles 10\% and 30\% on the
blue wing, and 70\%, 90\%, and 95\% on the red wing. In each of these wings we
evaluate $R_{\rm[N~II]/[S~II]}$. This is reported in the same figure and
associated with the corresponding velocity: the ratio increases from the value
$\sim 3$ measured in the Palomar spectrum, to $\sim 3.5$ for the whole line in
the HST observations, and consistently grows on both wings to $\sim 9$.
 
We repeated this analysis for all LINERs of the sample, by also deriving
  a radius at which the wing emission is produced, based on the assumption of
  Keplerian rotation. In Fig. \ref{wings}, lower panel, we plot the median
  ratios obtained in each wing (normalized to the ground-based value) and the
  corresponding median radius. The radial distances reach $\sim$1 pc for the
  5\% and 95\% percentiles, and we observe an increase of the ratio by a factor
  $\sim$ 5 times with respect to the value measured from the ground.
  
\landscape
\begin{table}
\caption{Forbidden line measurements}
\begin{tabular}{l|c|c|c|c|c|c|c|c|c|c|c|c|c|c|c|c|c}
\hline\hline
Name & \multicolumn{7}{|c|}{[SII]}  
     & \multicolumn{7}{|c|}{[NII]} 
     & \multicolumn{2}{|c|}{[OI]}  \\
\hline
     & \multicolumn{4}{|c|}{HST} & \multicolumn{3}{|c|}{Palomar} 
     & \multicolumn{4}{|c|}{HST} & \multicolumn{3}{|c|}{Palomar} 
     & \multicolumn{2}{|c|}{HST} \\
\hline
     & Flux & FWHM & v90-v10 & r
     & Flux & FWHM & r
     & Flux & FWHM & v90-v10 & r
     & Flux & FWHM & r
     & Flux & FWHM \\
\hline
\multicolumn{17}{l}{Liners Type 1}\\
\hline
3031 & -13.42 &    436  &    443  & 0.59 & -12.75 &    230 & 0.92  & -12.82 &    539  &    875  & 0.40 & -12.55 &    273  & 0.43 & -13.04 &    461  \\
3998 & -13.05 &    707  &   1626  & 0.79 & -12.75 &    439 & 1.14  & -12.75 &    968  &   1304  & 0.91 & -12.63 &    460  & 0.82 & -12.62 &   1326  \\
4203 & -13.83 &    256  &    434  & 0.75 & -13.58 &    371 & 1.01  & -13.29 &    441  &    768  & 0.90 & -13.36 &    372  & 0.75 & -13.28 &    592  \\
4450 & -13.69 &    368  &    491  & 0.84 & -13.38 &    311 & 1.01  & -13.61 &    416  &    630  & 0.73 & -13.51 &    312  & 0.47 & -13.49 &    653  \\
4579 & -13.80 &    536  &    848  & 0.74 & -12.92 &    388 & 1.10  & -13.61 &    565  &    701  & 0.30 & -12.84 &    390  & 0.51 & -13.54 &    980  \\
\hline
\multicolumn{17}{l}{Liners Type 2}\\
\hline 
 315 & -14.46 &    611  &    805  & 0.73 & -13.96 &    386 & 1.22  & -13.90 &    858  &   1377  & 0.31 & -13.56 &    436  & 0.43 &     -- &      -- \\
1052 & -13.44 &    572  &    739  & 0.74 & -12.18 &    509 & 1.13  & -13.31 &    646  &   1579  & 1.29 & -12.20 &    618  & 0.93 & -13.31 &    637  \\
1961 & -14.65 &    418  &    752  & 0.88 & -13.69 &    386 & 1.22  & -14.45 &    493  &    766  & 0.40 & -13.45 &    384  & 0.59 & -14.97 &    514  \\
2787 & -14.08 &    508  &    538  & 0.79 & -13.58 &    376 & 1.09  & -13.68 &    583  &   1255  & 0.44 & -13.35 &    403  & 0.68 &     -- &      -- \\
2911 & -15.49 &    371  &    417  & 1.23 & -13.33 &    485 & 1.28  & -15.05 &    704  &    999  & 0.34 & -13.26 &    499  & 0.59 &     -- &      -- \\
3642 & -14.62 &    182  &    470  & 0.89 & -13.61 &    189 & 1.29  & -14.37 &    323  &    800  & 0.99 & -13.67 &    178  & 1.98 &     -- &      -- \\
4036 & -14.64 &    494  &    538  & 0.89 & -13.24 &    381 & 1.22  & -14.34 &    557  &    718  & 0.32 & -13.12 &    386  & 0.52 & -14.99 &    471  \\
4143 & -13.84 &    463  &    794  & 0.78 & -13.77 &    260 & 1.10  & -13.31 &    614  &   1366  & 0.42 & -13.32 &    360  & 0.86 & -13.84 &    669  \\
4258 & -13.92 &    401  &    500  & 0.78 & -13.39 &    324 & 1.05  & -13.52 &    708  &    912  & 1.63 & -13.21 &    348  & 1.39 & -13.78 &    459  \\
4278 & -14.62 &    780  &   1366  & 0.71 & -12.72 &    484 & 1.14  & -14.35 &    960  &   1824  & 0.62 & -12.70 &    506  & 0.88 & -14.69 &   1379  \\
4477 & -14.59 &    223  &    422  & 0.89 & -13.65 &    296 & 1.18  & -14.17 &    302  &    944  & 0.37 & -13.41 &    277  & 0.63 &     -- &      -- \\
4486 & -13.94 &   1547  &   1887  & 1.50 & -12.98 &   1139 & 1.50  & -13.47 &      -- &      -- & 0.28 & -12.74 &   1139  & 0.39 &     -- &      -- \\
4548 & -15.72 &    165  &    217  & 1.11 & -15.31 &     71 & 1.18  & -15.53 &    190  &    285  & 0.31 & -15.92 &    218  & 0.49 &     -- &      -- \\
5005 & -14.03 &    399  &    745  & 1.00 & -12.74 &    683 & 1.14  & -13.75 &    494  &    893  & 0.22 & -12.45 &    747  & 0.35 &     -- &      -- \\
5077 & -14.33 &    898  &   1279  & 0.69 & -13.54 &    492 & 1.29  & -14.18 &    871  &   1252  & 0.78 & -13.49 &    522  & 0.92 &     -- &      -- \\
5377 & -14.58 &    130  &    498  & 0.90 & -13.66 &    296 & 1.19  & -14.27 &    261  &   1397  & 0.23 & -13.51 &    313  & 0.38 &     -- &      -- \\
6500 & -14.49 &    933  &   1666  & 0.93 & -13.01 &    560 & 1.14  & -14.50 &    763  &   1220  & 1.22 & -13.12 &    548  & 1.37 &     -- &      -- \\
\hline
\multicolumn{17}{l}{Seyferts Type 1}\\
\hline
3227 & -13.69 &    272  &    464  & 0.70 & -12.49 &    499 & 0.94  & -13.37 &    257  &    336  & 1.24 & -12.24 &    446  & 0.58 & -13.84 &    233  \\
3516 & -15.09 &    246  &    238  & 0.62 & -13.20 &    311 & 0.93  & -14.17 &      -- &      -- & 4.21 & -12.76 &    312  & 1.05 & -15.59 &    188  \\
4051 & -14.82 &    187  &    304  & 0.71 & -12.96 &    235 & 0.98  & -14.14 &      -- &      -- & 6.63 & -12.59 &    235  & 2.04 &     -- &      -- \\
\hline
\multicolumn{17}{l}{Seyferts Type 2}\\
\hline
1358 & -14.08 &    395  &    373  & 0.93 & -13.17 &    256 & 1.15  & -13.74 &    441  &    463  & 0.43 & -12.87 &    271  & 0.55 & -14.46 &    417  \\
1667 & -14.33 &    330  &    479  & 0.80 & -13.12 &    367 & 1.03  & -13.87 &    332  &    595  & 0.44 & -12.74 &    377  & 0.44 &     -- &      -- \\
2273 & -13.71 &    148  &    306  & 0.70 & -12.92 &    206 & 0.99  & -13.41 &    120  &    345  & 0.74 & -12.68 &    205  & 1.26 & -13.98 &    156  \\
3982 & -15.25 &    267  &    438  & 0.83 & -13.58 &    236 & 0.83  & -15.03 &    322  &    523  & 0.91 & -13.38 &    236  & 1.26 & -15.21 &    280  \\
4501 & -14.42 &    196  &    359  & 1.06 & -13.64 &    205 & 1.04  & -14.00 &    230  &    400  & 0.38 & -13.26 &    192  & 0.50 &     -- &      -- \\
4698 & -14.64 &    227  &    467  & 1.01 & -13.85 &    208 & 1.12  & -14.39 &    252  &    486  & 0.72 & -13.73 &    213  & 0.84 &     -- &      -- \\
5194 & -14.69 &    146  &    162  & 0.86 & -13.07 &    227 & 0.96  & -14.12 &    157  &    271  & 0.38 & -12.51 &    243  & 0.36 & -15.22 &    154  \\
6951 & -14.57 &    190  &    206  & 0.82 & -13.57 &    248 & 0.98  & -14.10 &    159  &    215  & 0.22 & -13.16 &    243  & 0.40 &     -- &      -- \\
\hline
\hline
\end{tabular}
\label{bigtable}
\medskip
\small{ 
\caption{(1) Object name  
From HST spectra:  (2) \sii flux (3) FWHM(\sii) (4) ratio of the \sii doublet.
From Palomar spectra: (5) \sii flux (6) FWHM(\sii) (7) ratio of the \sii doublet.
From HST spectra:  (8) \nii flux (9) FWHM(\nii) (10) ratio of the \nii doublet  
From Palomar spectra: (11) \nii flux (12) FWHM(\nii) (13) ratio of the \nii doublet.
From HST spectra:  (14) \oi flux (15) FWHM(\oi)  
From Palomar spectra: (16) \oi flux (17) FWHM(\oi). Fluxes are in logarithmic units of erg cm$^{-2}$ s$^{-1}$ and the FWHM in units of km s$^{-1}$.}
 }
\end{table}
\medskip
\endlandscape 

\section{The properties of the INLR in LINERs}

We summarize the main results obtained so far for the LINERs:
\begin{itemize}

\item the BH sphere of influence is resolved at the HST resolution in
  almost all galaxies;

\item the velocity dispersion of the \nii\ line correlates with the stellar
  velocity dispersion: this supports the idea that the INLR
  motions are dominated by the BH potential;

\item with the virial assumption, we can convert the FWHM velocity to a radial
  distance from the BH and we obtain a range of characteristic radii of the
  INLR between 1 and 10 pc, with a median of $\sim$3 pc.

\item the radius of the INLR is $\sim 4$ times the dust sublimation radius;

\item in the Palomar spectra, the velocity dispersion of the \nii\ does not
  differ from that of the \sii. In the HST spectra, the FWHM of the \nii\ line
  is significantly larger with respect to the FWHM of \sii, particularly when
  considering the wings of the lines. The same broadening effect is seen
  considering the width of the \oi\ line with respect to \sii;

\item the ratio of the two \sii\ lines decreases from Palomar to HST spectra,
  indicating an increase in the gas density;

\item the ratio between the \nii\ and the \sii\ increases from Palomar to HST
  spectra; we observe a further increase in the wings of the lines (probing a
  region extending over just one pc) with respect to the whole line in the HST
  spectra.

\end{itemize}

The above description is valid for most, but not all, LINERs. If we consider the
three main properties, which are broadening of the \nii, presence of broader \nii\ wings
in the HST spectra, and increase in the \nii/\sii\ ratio, we find only one
consistent exception. This exception is  NGC~1961, which is a source in which the BH sphere
of influence is not resolved. Another source (NGC~5005) fulfills only one of
these main properties.

The superior HST spatial resolution enables us to probe the innermost part of
the narrow line region that in ground-based spectra is swamped by emission
produced from gas located out to $\sim$ 100 pc. The changes in line widths and
ratios indicate that the physical conditions of the INLR gas in the parsec
region differ from what is seen at larger scales.

In particular, we noticed differences in the widths and intensity ratio between
\nii\ and \sii. Their ratio depends on many factors, i.e., the ISM metallicity,
the electron density of the emitting gas, and the properties of the ionizing
continuum. No strong dependence of the gas ionization level or temperature is
instead expected since the ionization potentials for \sii\ and \nii\ are
similar.\footnote{The ionization potentials are 23.33, 13.6, and 29.6 eV, for
  \sii, \oi\ and \nii, respectively \citep{moore70}. The logarithms of
  critical densities, in cm$^{-3}$ units, are ~3.2, 3.6, 6.3, and 4.9 for [S
    II]$\lambda$6716, [S II]$\lambda$6731, [O I]$\lambda$6300, and [N
    II]$\lambda$6584, respectively (\citealt{appenzeller88}).} When
considering individual sources, the dominant parameter is then the gas density
since the \sii\ doublet emission has a much lower critical density with
respect to \nii. The presence of regions where the density exceeds the
critical density of \sii\ produces an increase of the \nii/\sii\ intensity
ratio. This effect becomes more prominent where the gas velocity is larger,
i.e., at the smallest radii on the order of one pc. This leads to larger
\nii\ widths and suppression of the \sii\ emission in the line wings.

The picture that emerges is that the INLR in LINERs is characterized by the
presence of a structure of $\sim 10^{4}$ solar masses of ionized gas located
at a typical radius of $\sim$ 3 pc, a distance at which the gravitational
potential is dominated by the central BH. The changes of line widths and line
ratios (when moving from the large scale sampled by the Palomar data to the
HST spectra, but also seen in the comparison between the lines cores and
wings) are all accounted for by a density stratification within the INLR. In
particular, this implies the presence of regions where the gas density largely
exceeds the critical density of the \sii\ lines and possibly reaches values as
high as the \nii critical density, i.e., $\sim 10^5$cm$^{-3}$. The
  density and gas velocity are intermediate between those measured in the NLR
  and in the BLR and this is reminiscent of the properties of the ILR,
  initially identified in UV spectra of Seyferts. Therefore, we suggest that
  our optical spectra are mapping the outer portion of the ILR.

\section{The properties of the INLR in Seyfert galaxies}

The situation for the Seyfert galaxies is radically different. We do not
observe any significant change in the lines widths and ratios from the Palomar
to the HST spectra, both for the lines as a whole and for the line wings. The
only variation is seen for the relative intensity of the \sii\ doublet,
indicative of a modest density increase from the large-scale NLR to the INLR
from a few hundreds to $\sim$ $10^3$cm$^{-3}$. The lack of variation in the
\sii\ width and ratio against \nii\ and \oi\ suggests that there is no
evidence for emission from gas in the INLR with a density intermediate
between the \sii\ and \oi\ critical density.

In the Seyferts even HST does not, in general, resolve the BH sphere of
influence. We found only three exceptions (namely, NGC~5194, NGC~4698, and
NGC~4501; see Fig. \ref{sphere}), and all of these are type 2 objects. We estimate
a radial distance of the emitting gas in the same way we have done for
LINERs. In Fig. \ref{torus} we compare $R_{\rm FWHM[NII]}$ with the dust
sublimation radius: these three Seyferts span a similar range in radial
distance from the BH (between 2 and 10 parsec) covered by the LINERs, and
larger than the dust sublimation radius, $R_{\rm d}$, by a factor 2 to
10. Since $R_{\rm d}$ can be associated with the inner radius of the
circumnuclear obscuring structure, the observed gas lies, most likely, above
(and below) the torus. As shown in Fig. \ref{torus}, this result also applies 
to the remaining Seyferts. 

We conclude that the pc scale structure of ionized (and stratified) gas seen
in LINERs, which we identify as the outer portion of the ILR, is not visible in
Seyfert galaxies. The gas seen in the HST spectra appears to be simply the
extension of the large-scale NLR to smaller radii, probably located at the
base of the ionization cones with densities reaching at most $\sim$
$10^3$cm$^{-3}$.

For Seyfert type 2, there is the possibility that the ILR might be obscured by
the torus.  Unfortunately, we have only three type 1 Seyferts with HST data and
in only one (NGC~3227), and we are able to deblend  the \Ha\, broad
emission from the \Ha+\nii complex reliably. Thus, we measure the flux and FWHM of
\nii. For another type 1 Seyfert (NGC~3516), we can only compare the width of
the \oi\ and \sii with no apparent broadening.  Although this analysis is
only limited to  two objects, both behave like the type 2 Seyfert. 

A possible explanation is that the emission originating from the outer
  portion of the ILR is diluted by the NLR emission. In fact, in Seyferts the
  HST slit maps a larger region and, therefore, the different behavior observed
  in LINERs and Seyferts can be ascribed to resolution effects.  Only the
  coronal lines of higher ionization coming from more dense, compact and,
  therefore, bright regions show the existence of the ILR. Conversely, the
  stratification effect is not evident in optical lines produced in the more
  tenuous and diffuse medium at larger scale.

\section{Discussion}
\label{discussion}

The topic of NLR stratification has been the subject of many studies (see, for
example, \citealt{whittle85} and references therein). In particular,
\citeauthor{whittle85} found no correlations between widths and critical
density or ionization potential in the five Seyfert 2 studied, which is in line with
our results. These correlations are instead found for Seyfert 1 and, in general,
the stronger trend is between the line width and its critical density,
apparently in contrast with our findings. However, this study takes
advantage of the analysis of a larger number of lines, whose critical
densities reach much higher values that those considered in our analysis with
values in excess of $\sim$ $10^7$cm$^{-3}$. The comparisons between the
\oi\ and \sii\ widths, possible for eleven sources, do not show any significant
differences (with only one exception, namely MRK~9), again, in agreement with
our results. The NLR density stratification is present but it is associated
with much denser gas than sampled by the lines present in the HST data we
considered. We can envisage that this occurs at very small radii, well inside
the obscuring torus.  The results presented by \citet{nagao01} support this
interpretation since they found difference in the properties of the
\oiii$\lambda4363$ line, having a critical density of 10$^{7.41}$ cm$^{-3}$,
between Sy 1 and Sy 2. They argue that this line, as well as other high
ionization lines, can be produced from the inner wall of the torus. Other
studies found that indeed high ionization emission lines are stronger in Sy1s
than in Sy2s (\citealt{murayama98}; 1998a; \citealt{nagao00},
\citealt{baskin05}, \citealt{baldi13}, \citealt{dicken14}).

\begin{figure}
\centering{
\includegraphics[scale=0.25,angle=0]{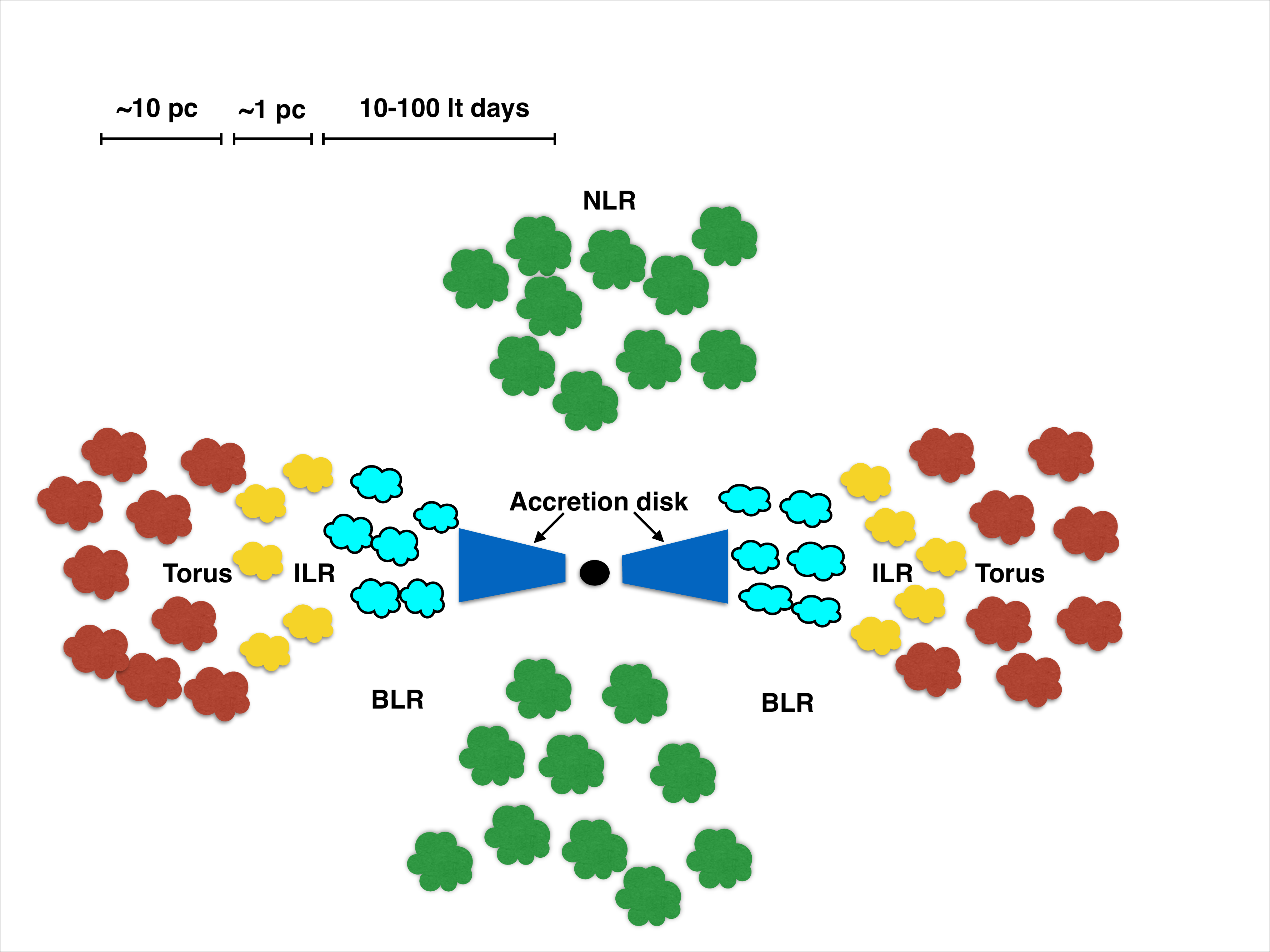}
\includegraphics[scale=0.25,angle=0]{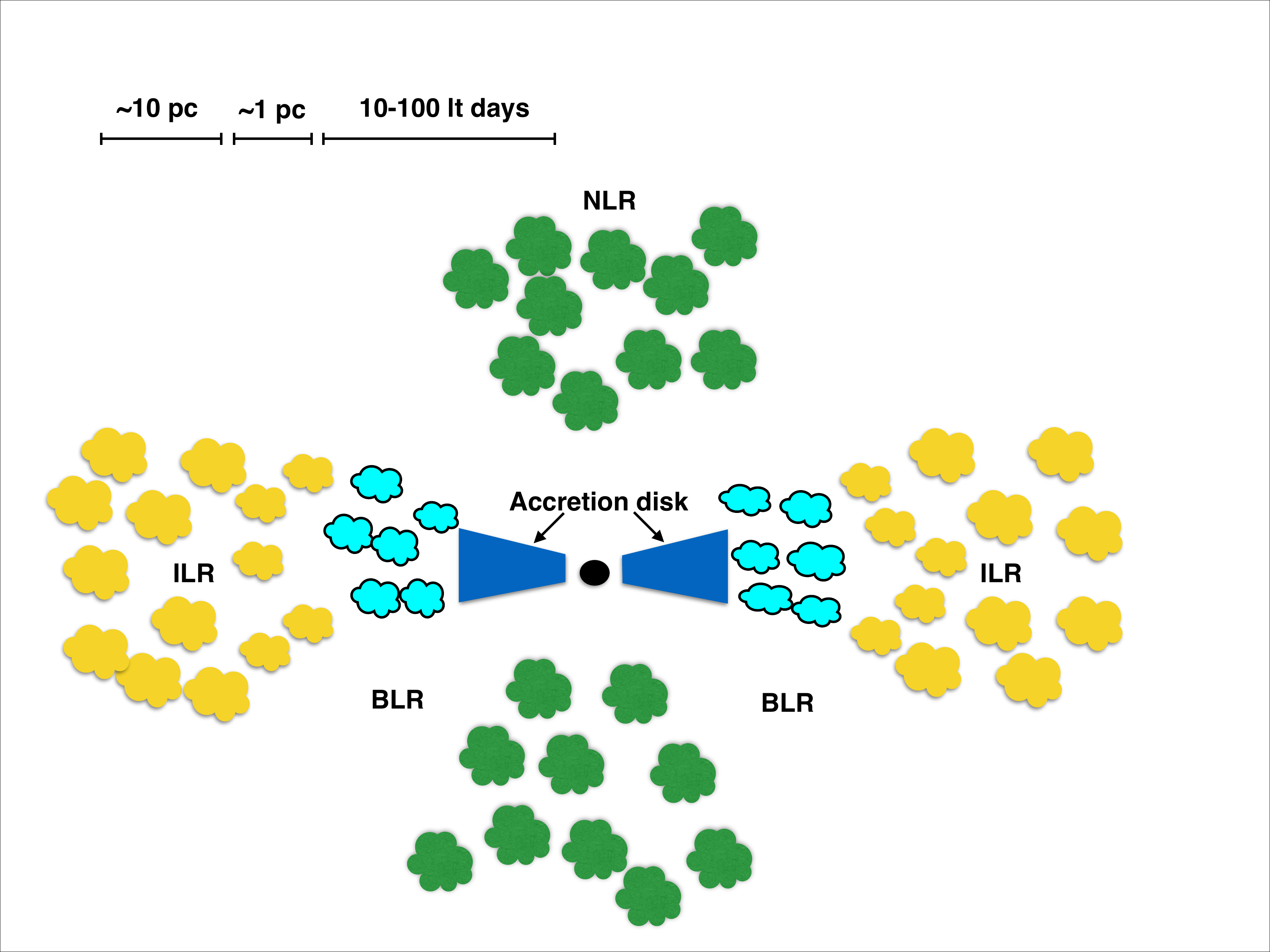}}
\caption{Schematic view of the ILR in Seyfert galaxies (top panel) and in LINER galaxies (bottom panel). The image is not on scale.}
\label{draw}
\end{figure}

Our results on the NLR properties of low-luminosity AGN are accounted for by
considering the interplay between the line emitting regions and the obscuring
torus (see Fig. \ref{draw} for a schematic view). A toroidal absorbing
structure is required in Seyferts (e.g., \citealt{antonucci93}) to explain the
differences between type 1 and type 2 objects. Seyfert 2 galaxies are indeed
characterized by high values of hydrogen column density and about half of
 nearby Seyfert 2 are Compton thick, i.e., have $N_{\rm H}$ $>$ 10$^{24}$
cm$^{-2}$ \citep{guainazzi05}. Conversely, several pieces of evidence point
to the conclusion that the obscuring torus is in general not present in
LINERs. In particular, the main effect associated with the torus, i.e., the
absorption in the X-ray spectra, is not visible in this class of AGN
\citep{balmaverde15} with $N_{\rm H}$ rarely exceeding $\sim$ 10$^{22}$
cm$^{-2}$.

The overall structure of the emission line region in LINERs can be described
as a gas distribution in which the density smoothly decreases moving outward.
In Seyfert galaxies, the circumnuclear gas becomes optically thick for the
  UV ionizing photons, thus, forming the obscuring torus.  The molecular torus
produces a gap in the observed gas density distribution, occurring on a scale
of $\sim$ 3 pc and corresponding to densities in the range 10$^4$ - 10$^6$
cm$^{-3}$. At even smaller radii, within the dust sublimation radius, the
density further increases leading to the known link between critical density
and line width.

The results obtained for the LINERs are reminiscent of what was found by
\citet{capetti05} analyzing HST narrowband images in low-luminosity FR~I
radio galaxies. These authors similarly identify a compact emission line
region extending on pc-scale. In line with their interpretation, we suggest
that the ILR takes the form of an ionized, optically thin torus. We suggest
that this tenuous structure is present only in LINERs because of the general
paucity of gas and dust in their nuclear regions. This also causes their low
rate of accretion and low bolometric luminosity. A standard torus can only
form in more powerful AGN
associated with larger amounts of nuclear gas and dust.

\section{Summary and conclusions}
\label{summary}

The aim of this paper is to compare the properties of the optical emission
lines of the NLR as seen in HST/STIS and in ground-based spectra in low-luminosity AGN. The superior spatial resolution of HST offers the possibility
of exploring the properties of the innermost regions of the NLR with respect
to what is seen at larger scales.

We consider 33 Seyferts and LINERs galaxies from the Palomar sample observed
by HST/STIS with the high-resolution grism centered to cover the \Ha\,
emission.  We compare the width and  intensity ratio between different
emission lines in Palomar and HST spectra.

In LINERs, we find that the FWHM of the \nii and the ratio between the
intensity of \nii and \sii increase from Palomar to HST spectra, and it is
even higher in the wings of the lines (defined at different percentiles of the
flux).  This could be interpreted as evidence that the gas density increases
toward the center, exceeding the \sii critical density.  Assuming that the
kinematic is dominated by the BH gravitational potential, we convert the
velocities to a radial distance from the BH and we find radii on the order of
$\sim$1-10 pc.  We identify this compact, high density region, as the outer portion of the intermediate line
region or ILR.
 
This region is instead invisible in Seyferts, where the line widths and ratios
do not change significantly from the ground-based to HST data and do not
differ in the different lines.
  
The absorption in the X-ray data indicates that the standard torus is present
in Seyferts, while it is generally absent in LINERs. Therefore, the most
plausible scenario is that in LINERs the ILR takes the place of the
circumnuclear torus. We suggest that the outer ILR, the region of optically
thin, ionized gas, and the optically thick torus are mutually exclusive
structure.  A standard torus is only present in powerful AGN associated with larger amounts of nuclear gas
and dust.  Instead, in LINERs, as a result of the general
paucity of interstellar material in their nuclear regions, only a tenuous,
ionized, and optically thin structure can form.

The different gas structure in Seyfert and LINERs indicates that the latter
class is a superior target for BH mass measurements based on gas
dynamics.  The presence of the ILR, in fact, represents the ideal component for
such a study, as it extends well within the BH sphere of influence and it
contains a large fraction of the total line emission.  Conversely, in Seyferts,
most of the emission from the lines commonly used is produced outside the
obscuring torus. Our results suggest that, for Seyferts, lines of higher
critical density (such as, for example, [Ne~III]$\lambda3869$ and
[Ne~V]$\lambda\lambda3346,3426$) that are emitted at smaller radii should be
considered.

\begin{acknowledgements}
This research has made use of the NASA/IPAC Extragalactic Database (NED), which
is operated by the Jet Propulsion Laboratory, California Institute of
Technology, under contract with the National Aeronautics and Space
Administration. We thank the referee for her/his useful suggestions and comments.
\end{acknowledgements}

\bibliographystyle{aa} 


\begin{thebibliography}{37}
\expandafter\ifx\csname natexlab\endcsname\relax\def\natexlab#1{#1}\fi

\bibitem[{{Antonucci}(1993)}]{antonucci93}
{Antonucci}, R. 1993, \araa, 31, 473

\bibitem[{{Appenzeller} \& {Oestreicher}(1988)}]{appenzeller88}
{Appenzeller}, I. \& {Oestreicher}, R. 1988, \aj, 95, 45

\bibitem[{{Baldi} {et~al.}(2013){Baldi}, {Capetti}, {Buttiglione}, {Chiaberge},
  \& {Celotti}}]{baldi13}
{Baldi}, R.~D., {Capetti}, A., {Buttiglione}, S., {Chiaberge}, M., \&
  {Celotti}, A. 2013, \aap, 560, A81

\bibitem[{{Balmaverde} \& {Capetti}(2014)}]{balmaverde14}
{Balmaverde}, B. \& {Capetti}, A. 2014, \aap, 563, A119

\bibitem[{{Balmaverde} \& {Capetti}(2015)}]{balmaverde15}
---. 2015, ArXiv e-prints

\bibitem[{{Baskin} \& {Laor}(2005)}]{baskin05}
{Baskin}, A. \& {Laor}, A. 2005, \mnras, 358, 1043

\bibitem[{{Blandford} {et~al.}(1990){Blandford}, {Netzer}, {Woltjer},
  {Courvoisier}, \& {Mayor}}]{netzer90}
{Blandford}, R.~D., {Netzer}, H., {Woltjer}, L., {Courvoisier}, T.~J.-L., \&
  {Mayor}, M., eds. 1990, {Active Galactic Nuclei}

\bibitem[{{Capetti} {et~al.}(2005){Capetti}, {Verdoes Kleijn}, \&
  {Chiaberge}}]{capetti05}
{Capetti}, A., {Verdoes Kleijn}, G., \& {Chiaberge}, M. 2005, \aap, 439, 935

\bibitem[{{Crenshaw} \& {Kraemer}(2007)}]{crenshaw07}
{Crenshaw}, D.~M. \& {Kraemer}, S.~B. 2007, \apj, 659, 250

\bibitem[{{Crenshaw} {et~al.}(2009){Crenshaw}, {Kraemer}, {Schmitt}, {Kaastra},
  {Arav}, {Gabel}, \& {Korista}}]{crenshaw09}
{Crenshaw}, D.~M., {Kraemer}, S.~B., {Schmitt}, H.~R., {et~al.} 2009, \apj,
  698, 281

\bibitem[{{Dicken} {et~al.}(2014){Dicken}, {Tadhunter}, {Morganti}, {Axon},
  {Robinson}, {Magagnoli}, {Kharb}, {Ramos Almeida}, {Mingo}, {Hardcastle},
  {Nesvadba}, {Singh}, {Kouwenhoven}, {Rose}, {Spoon}, {Inskip}, \&
  {Holt}}]{dicken14}
{Dicken}, D., {Tadhunter}, C., {Morganti}, R., {et~al.} 2014, \apj, 788, 98

\bibitem[{{Elitzur}(2008)}]{elitzur08}
{Elitzur}, M. 2008, \nar, 52, 274

\bibitem[{{Filippenko} \& {Sargent}(1985)}]{filippenko85}
{Filippenko}, A.~V. \& {Sargent}, W.~L.~W. 1985, \apjs, 57, 503

\bibitem[{{Gaskell}(1984)}]{gaskell84}
{Gaskell}, C.~M. 1984, \aplett, 24, 43

\bibitem[{{Guainazzi} {et~al.}(2005){Guainazzi}, {Matt}, \&
  {Perola}}]{guainazzi05}
{Guainazzi}, M., {Matt}, G., \& {Perola}, G.~C. 2005, \aap, 444, 119

\bibitem[{{Heckman} {et~al.}(2004){Heckman}, {Kauffmann}, {Brinchmann},
  {Charlot}, {Tremonti}, \& {White}}]{heckman04}
{Heckman}, T.~M., {Kauffmann}, G., {Brinchmann}, J., {et~al.} 2004, \apj, 613,
  109

\bibitem[{{Ho} {et~al.}(1995){Ho}, {Filippenko}, \& {Sargent}}]{ho95}
{Ho}, L.~C., {Filippenko}, A.~V., \& {Sargent}, W.~L. 1995, \apjs, 98, 477

\bibitem[{{Ho} {et~al.}(1997){Ho}, {Filippenko}, \& {Sargent}}]{ho97}
{Ho}, L.~C., {Filippenko}, A.~V., \& {Sargent}, W.~L.~W. 1997, ApJS, 112, 315

\bibitem[{{Ho} {et~al.}(2009){Ho}, {Greene}, {Filippenko}, \&
  {Sargent}}]{ho09a}
{Ho}, L.~C., {Greene}, J.~E., {Filippenko}, A.~V., \& {Sargent}, W.~L.~W. 2009,
  \apjs, 183, 1

\bibitem[{{Humphrey} {et~al.}(2008){Humphrey}, {Villar-Mart{\'{\i}}n},
  {Vernet}, {Fosbury}, {di Serego Alighieri}, \& {Binette}}]{humphrey08}
{Humphrey}, A., {Villar-Mart{\'{\i}}n}, M., {Vernet}, J., {et~al.} 2008,
  \mnras, 383, 11

\bibitem[{{Kraemer} \& {Crenshaw}(2000)}]{kraemer00}
{Kraemer}, S.~B. \& {Crenshaw}, D.~M. 2000, \apj, 532, 256

\bibitem[{{Marconi} {et~al.}(1996){Marconi}, {van der Werf}, {Moorwood}, \&
  {Oliva}}]{marconi96}
{Marconi}, A., {van der Werf}, P.~P., {Moorwood}, A.~F.~M., \& {Oliva}, E.
  1996, \aap, 315, 335

\bibitem[{{Moore}(1970)}]{moore70}
{Moore}, C.~E. 1970, {Ionization potentials and ionization limits derived from
  the analyses of optical spectra}

\bibitem[{{Moore} \& {Cohen}(1994)}]{moore94}
{Moore}, D. \& {Cohen}, R.~D. 1994, \apj, 433, 602

\bibitem[{{Moore} {et~al.}(1996){Moore}, {Cohen}, \& {Marcy}}]{moore96}
{Moore}, D., {Cohen}, R.~D., \& {Marcy}, G.~W. 1996, \apj, 470, 280

\bibitem[{{Murayama} \& {Taniguchi}(1998)}]{murayama98}
{Murayama}, T. \& {Taniguchi}, Y. 1998, \apjl, 503, L115

\bibitem[{{Nagao} {et~al.}(2006){Nagao}, {Marconi}, \& {Maiolino}}]{nagao06}
{Nagao}, T., {Marconi}, A., \& {Maiolino}, R. 2006, \aap, 447, 157

\bibitem[{{Nagao} {et~al.}(2001){Nagao}, {Murayama}, \& {Taniguchi}}]{nagao01}
{Nagao}, T., {Murayama}, T., \& {Taniguchi}, Y. 2001, \apj, 546, 744

\bibitem[{{Nagao} {et~al.}(2000){Nagao}, {Murayama}, {Taniguchi}, \&
  {Yoshida}}]{nagao00}
{Nagao}, T., {Murayama}, T., {Taniguchi}, Y., \& {Yoshida}, M. 2000, \aj, 119,
  620

\bibitem[{{Onken} {et~al.}(2004){Onken}, {Ferrarese}, {Merritt}, {Peterson},
  {Pogge}, {Vestergaard}, \& {Wandel}}]{onken04}
{Onken}, C.~A., {Ferrarese}, L., {Merritt}, D., {et~al.} 2004, \apj, 615, 645

\bibitem[{{Osterbrock}(1989)}]{osterbrock89}
{Osterbrock}, D.~E. 1989, {Astrophysics of gaseous nebulae and active galactic
  nuclei}

\bibitem[{{Peterson} {et~al.}(2004){Peterson}, {Ferrarese}, {Gilbert}, {Kaspi},
  {Malkan}, {Maoz}, {Merritt}, {Netzer}, {Onken}, {Pogge}, {Vestergaard}, \&
  {Wandel}}]{peterson04}
{Peterson}, B.~M., {Ferrarese}, L., {Gilbert}, K.~M., {et~al.} 2004, \apj, 613,
  682

\bibitem[{{Sarzi} {et~al.}(2006){Sarzi}, {Falc{\'o}n-Barroso}, {Davies},
  {Bacon}, {Bureau}, {Cappellari}, {de Zeeuw}, {Emsellem}, {Fathi},
  {Krajnovi{\'c}}, {Kuntschner}, {McDermid}, \& {Peletier}}]{sarzi06}
{Sarzi}, M., {Falc{\'o}n-Barroso}, J., {Davies}, R.~L., {et~al.} 2006, \mnras,
  366, 1151

\bibitem[{{Tremaine} {et~al.}(2002){Tremaine}, {Gebhardt}, {Bender}, {Bower},
  {Dressler}, {Faber}, {Filippenko}, {Green}, {Grillmair}, {Ho}, {Kormendy},
  {Lauer}, {Magorrian}, {Pinkney}, \& {Richstone}}]{tremaine02}
{Tremaine}, S., {Gebhardt}, K., {Bender}, R., {et~al.} 2002, \apj, 574, 740

\bibitem[{{Veilleux}(1991)}]{veilleux91}
{Veilleux}, S. 1991, \apj, 369, 331

\bibitem[{{Veilleux} \& {Osterbrock}(1987)}]{veilleux87}
{Veilleux}, S. \& {Osterbrock}, D.~E. 1987, \apjs, 63, 295

\bibitem[{{Whittle}(1985)}]{whittle85}
{Whittle}, M. 1985, \mnras, 216, 817

\end{thebibliography}
\end{document}